\documentclass[prx,nobibnotes,nofootinbib,twocolumn,showpacs]{revtex4-2}
\usepackage{mathrsfs}
\usepackage{calc}
\usepackage[T1]{fontenc}
\usepackage[latin1]{inputenc}
\usepackage{amsmath}
\usepackage{amssymb}
\usepackage{xspace}
\usepackage{graphicx}
\usepackage[normalem]{ulem}
\usepackage{xcolor}
\usepackage{multirow}

\usepackage{tikz}
\usetikzlibrary{arrows}
\usetikzlibrary{calc,decorations.pathmorphing,shapes}

\allowdisplaybreaks[1]

\newcommand{\be}{\begin{equation}}
\newcommand{\ee}{\end{equation}}

\newcommand{\p}{\partial}

\newcommand{\cM}{\mathcal{M}}
\newcommand{\cO}{\mathcal{O}}

\definecolor{darkgreen}{rgb}{0.2,0.6,0}
\definecolor{lightblue}{rgb}{0,0.5,0.8}
\definecolor{lightred}{rgb}{0.8,0.2,0.2}
\definecolor{darkorange}{rgb}{1,0.549,0}
\definecolor{brown}{rgb}{0.609, 0.164, 0.164}

\makeatother

\begin{document}

\definecolor{rvwvcq}{rgb}{0.08235294117647059,0.396078431372549,0.7529411764705882}
\definecolor{wrwrwr}{rgb}{0.3803921568627451,0.3803921568627451,0.3803921568627451}

\title{Regular evaporating black holes with stable cores}
\author{Alfio Bonanno}
\affiliation{INAF, Osservatorio Astrofisico di Catania, via S.Sofia 78 and INFN, Sezione di Catania, via S. Sofia 64, I-95123,Catania, Italy}
\author{Amir-Pouyan Khosravi}
\author{Frank Saueressig}
	\affiliation{Institute for Mathematics, Astrophysics and Particle Physics (IMAPP), \\ Radboud University, Heyendaalseweg 135, 6525 AJ Nijmegen, The Netherlands}
\pacs{04.60.-m, 11.10.Hi}

\begin{abstract}
A feature shared by many regular black hole spacetimes is the occurrence of a Cauchy horizon. It is then commonly believed that this renders the geometry unstable against perturbations through the mass-inflation effect. In this work, we perform the first dynamical study of this effect taking into account the mass-loss of the black hole due to Hawking radiation. It is shown that the time-dependence of the background leads to two novel types of late-time behavior whose properties are entirely determined by the Hawking flux. The first class of attractor-behavior is operative for regular black holes of the Hayward and renormalization group improved type and characterized by the square of the Weyl curvature growing as $v^6$ at asymptotically late times. This singularity is inaccessible to a radially free-falling observer though. The second class is realized by Reissner-Nordstr{\"o}m black holes and regular black holes of the Bardeen type. In this case the curvature scalars remain finite as $v\rightarrow\infty$. Thus the Hawking flux has a profound effect on the mass-inflation instability, either weakening the effect significantly or even expelling it entirely.
\end{abstract} 
\maketitle
\section{Introduction}
The resolution of spacetime singularities, omnipresent in solutions of general relativity \cite{Hawking:1970zqf}, constitutes a key motivation for quantum gravity research. In particular, singularity resolution would provide an elegant solution to the information loss problem in black hole physics \cite{Harlow:2014yka,Polchinski:2016hrw} since it would invalidate the conclusion that information falling into a black hole is inevitably destroyed when it hits the classical curvature singularity. While first principle investigations of the mechanisms which could lead to singularity resolution are still scarce \cite{UV-grav1,UV-grav2,UV-grav3,UV-grav4,Bosma:2019aiu}, there have been a number of phenomenologically motivated proposals for regular black hole geometries replacing the singular region by a patch of regular de Sitter space. Examples include the regular black hole geometries proposed by Dymnikova \cite{Dymnikova,Dymnikova1}, Bardeen \cite{bardeen1968non}, renormalization group improved black hole spacetimes in the context of the gravitational asymptotic safety program \cite{Bonanno:2000ep} (also see \cite{Koch:2014cqa,Platania:2019kyx,Scardigli:2022jtt} for recent developments and more references), Planck stars motivated from Loop Quantum Gravity \cite{Planck,Planck1,Saueressig:2015xua}, the Hayward black hole \cite{Hayward:2005gi}, and extensions proposed by Koshelev et.\ al.\ \cite{Koshelev-ext}.\footnote{Also see \cite{Frolov-rev} for a review and further references.}  More recently, regular black hole spacetimes exhibiting a ``Gauss'' core have been proposed in \cite{Boos:2021kqe}.

Based on topological considerations, it is clear that any regular black hole solution exhibiting an asymptotically flat region and a de Sitter core must have an even number of horizons. In the simplest case, this entails that there is an (outer) event horizon and an (inner) Cauchy horizon. This horizon structure is identical to the one of a charged Reissner-Nordstr{\"o}m black hole.  

This modification entails a drastic consequence for the final state of the black hole. When treated at the level of quantum field theory on a curved spacetime, black holes emit Hawking radiation, a perfect black body radiation whose temperature $T$ is proportional to the surface gravity of the event horizon. As the Hawking radiation carries away energy, the black hole becomes lighter and $T$ increases. For geometries with a single event horizon this process terminates with the complete evaporation of the black hole within a finite time. The presence of a inner horizon changes this picture drastically: in this case the energy loss due to Hawking radiation leads to the two horizons approaching each other. The final state of the evaporation process is then a critical black hole remnant with a typical mass given by the Planck mass \cite{Saueressig:2021pzy}. Reaching this final configuration requires an infinite time-span though.  

A central point of critique, challenging the phenomenological viability of such regular black hole solutions \cite{Frolov-rev1,Carballo-Rubio:2018pmi}, originates from the mass-inflation effect, first investigated in the context of the Reissner-Nordstr{\"o}m solution \cite{PI1990}. In short, the mass-inflation scenario considers a perturbation of the black hole geometry by outgoing null-radiation modeling, e.g., gravitational waves created in the black hole formation process and subsequently reflected by the potential barrier surrounding the black hole. These perturbations lead to an exponential growth of the mass function at the Cauchy horizon, ultimately generating a null singularity. While this scenario potentially resolves the ambiguities arising from extending geodesics beyond a Cauchy horizon \cite{PI1990,Ori:1991zz}, it also suggests that the geometries are not stable on the timescales associated with the black hole evaporation process. Extrapolating this scenario from the Reissner-Nordstr{\"om} geometry to regular black hole solutions then suggests that these could also suffer from the mass-inflation effect, leading to a dynamically generated spacetime singularity at the Cauchy horizon. 

In \cite{Bonanno:2020fgp}, it was established that this conclusion is premature though. While the extrapolation works for certain classes of \emph{static} regular black holes, including the geometries proposed by Bardeen, the Hayward geometry and renormalization group improved black hole solutions are free from mass-inflation. In these cases the mass function at the Cauchy horizon grows polynomially in time only and the resulting curvature singularity may be integrable. Technically, this behavior can be traced back to the presence of a late-time attractor in the evolution equation for the mass-function at the Cauchy horizon, rendering this quantity finite at asymptotically late times. 

 The taming of the mass-inflation effect then suggests that the dynamics at the Cauchy horizon and the black hole evaporation process could happen on similar timescales. Thus, a more complete understanding of the actual dynamics mandates to take the Hawking evaporation process into account. Our work addresses this question for the first time. As our main result, we discover two classes of universal late-time behaviors whose properties are dictated by simple structural properties of the mass function and the universality of the Hawking effect. The late-time attractors governing the dynamics either lead to a polynomial growth of the squared curvature tensors or even renders these quantities finite. Interestingly, the latter behavior appears for the case of Reissner-Nordstr{\"o}m geometry once the Hawking radiation is included. The final state of the black hole evaporation process is still a cold remnant.

The rest of our work is then organized as follows. The Ori-model for mass-inflation \cite{Ori:1991zz} is reviewed in Sect.\ \ref{sect.2}. Sects.\ \ref{sect.3} and \ref{sect.4} contains a detailed discussion of the model in the context of an evaporating Hayward black hole and the Reissner-Nordstr{\"o}m geometry, respectively. Other regular black hole geometries can be treated along similar lines and are discussed in Sect.\ \ref{sect.RBH}. Our conclusions and a brief outlook are given in Sect.\ \ref{sect.6}. Technical details related to our analysis are given in three appendices with Appendix \ref{app.A} summarizing the explicit expressions for the scalar curvature invariants, Appendix \ref{App.B} giving the detailed results for the late-time attractors, and Appendix \ref{app.C} analyzing the properties of radial geodesics in the black hole geometries.

\section{The Ori-model for mass inflation}
\label{sect.2}

\begin{figure}
\centering
\resizebox{0.4\textwidth}{!}{\begin{tikzpicture}
\tikzset{->-/.style={decoration={markings,mark=at position {#1} with {\arrow{>}}}, postaction={decorate}}}
\tikzset{-<-/.style={decoration={markings,mark= at position {#1} with {\arrow{<}}},
postaction={decorate}}}
\draw[white,ultra thin] (-3,-3) grid (3,3);
\draw (-2+2,3-2) -- (2+1,-1-1) node[pos=.6,right] {$\;\mathcal{J}^+$} node[pos=.15,sloped,above] {$f_{-}=0$} 
node[pos=.15, sloped, below] {$v=\infty$};
\draw [dashed,thick] (-2,3) -- (0,1) node[pos=0.2, sloped, below] {CH} ;
\draw (-2+0.5,3+0.5) -- (0+0.5,1+0.5) node [pos=0.4, sloped,above] {$f_{+}=0$};
\draw [thick] (1,0) to [out=-135,in=40] (1,-1.5);
\draw [thick] (1,-1.5) -- (0,-2.5) node[pos=0.6,sloped,below] {AH};
\node at (-1.5,0.5) {${\cal M_{+}}$};
\node at (-0.5,-0.5) {${\cal M_{-}}$};
\draw [dashed] (-3.5+2,-4.5+2) -- (1,0) node[pos=0.2,sloped,below] {EH};
\draw [thick] (-3.5+1,-2.5+1) -- (0.5,1.5) node[pos=.2,sloped,below] {$\Sigma$};
\draw [->, decoration={snake, amplitude=0.5mm, segment length = 2mm}, decorate,thick]  
(2, -1.5 ) -- (0.5,-0) node[pos=-0.4,sloped] {$v=const$};
\end{tikzpicture}}
	\caption{Schematic late advanced-times conformal geometry of our model. The event horizon (EH) and Cauchy horizon (CH) are indicated by the dashed lines while the position of the (time-dependent) apparent horizon (AH) has been added as the solid curve. Owed to the Hawking flux (indicated by the arrow) these quantities agree at asymptotically late times $v=\infty$. The shell $\Sigma$ separates spacetime into an inner region $\cM_+$ and an outer region $\cM_-$. The spherically symmetric geometries in each sector are characterized by the lapse functions $f_\pm$, respectively. \label{fig.conformal}}
\end{figure}
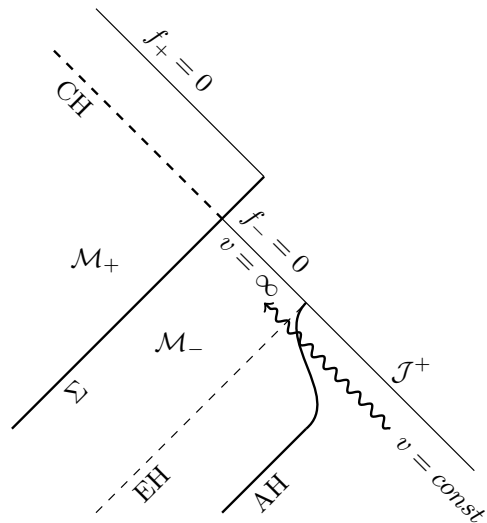

We start by reviewing the background material underlying the analysis in the main parts of our work. 
The line-element of a (generically non-static) spherically symmetric spacetime can be cast into a $2+2$ form,
\begin{equation}
\label{sphericalsymline}
ds^2= g_{ab}(x^c) dx^a dx^b+r^2 d\Omega^2 \, , 
\end{equation}
where $d\Omega^2=d\theta^2+\sin^2\theta d\phi^2$ 
is the line-element on the unit two-sphere, $x^a$, $a=1,2$, are coordinates in the submanifold $\theta=\phi={\rm const}$,
and $r(x^c)$ is the radius of the 2-sphere $x^a=const$. A physically relevant quantity is the quasi-local Misner-Sharp mass function $M(x^c)$ defined via
\begin{equation}
\label{lapse}
g^{ab}\partial_a r\partial_b r = f(x^c) = : 1 -\frac{2 M(x^c)}{r} \, . 
\end{equation}
For a Schwarzschild black hole $M(x^c)$ agrees with the mass-function $m$. Static, regular black hole geometries typically generalize this relation by promoting $M$ to a function of the radial coordinate.


The mass-inflation phenomenon occurs when the inner horizon is perturbed by a cross-flow stream of light-like matter. In a realistic gravitational collapse this combination of outgoing and ingoing flux is produced by the matter of the collapsing star and by the combined contribution of ingoing and outgoing gravitational waves \cite{Hamilton:2008zz}.
In the optical geometric limit one can simply consider light-like pressureless matter and use a coordinate system adapted to the null-generators of the ingong radiation, so that 
$v=constant$ is an ingoing null ray. 
%
%
 Ori has shown that the physics of the instability can be discussed assuming that the outgoing  flux is modeled by 
  an infinitesimally thin, pressureless null shell $\Sigma$. Albeit this assumption is valid in the optical geometric limit only, the model has the advantage to be analytically tractable in the $v\rightarrow\infty$ limit. 
  
 The shell $\Sigma$ separates spacetime into two regions $\cM_{\pm}$ with $(+)$ and $(-)$ referring to the regions inside and outside the shell, respectively. The metric in each sector can then  be written as
\be\label{eq.regions}
ds^2 = -f_{\pm}(r,v_{\pm}) dv_\pm + 2 dr dv_\pm + r^2 d\Omega^2 \, . 
\ee
The lapse function $f_{\pm}(r,v_{\pm})$ in each sector depends on $r$ and the ingoing Eddington-Finkelstein coordinate $v_\pm$ according to \eqref{lapse}.
The equality of the induced metrics on $\Sigma$ forces the radial coordinate $r$ to be equal on both sides of the shell. The $v$-coordinates in the two regions are related by 
\be\label{fvrel}
f_+(r,v_+) \, dv_+ = f_-(r,v_-) \, dv_- \, , 
\ee 
along $\Sigma$. This relation allows to express $v_+$ in terms of $v_-$ and we chose to express the dynamics in terms of $v \equiv v_-$.

The position of the shell $R(v)$, as a function of $v$ is governed by the first order differential equation
\be\label{Rpos}
\dot{R}(v) = \left. \frac{1}{2} f_- \right|_{\Sigma} \, , 
\ee
where the dot represents a derivative with respect to $v$. Furthermore, the Misner-Sharp mass in the two sectors separated by the shell are related by \cite{Ori:1991zz,Bar-Isr}
\be
\left. \frac{1}{f_+^2} \frac{\p M_+}{\p v_+} \right|_\Sigma =
\left. \frac{1}{f_-^2} \frac{\p M_-}{\p v_-} \right|_\Sigma
\ee
with the $v$-derivatives evaluated before substituting the position of the shell. Using the identity \eqref{fvrel}, this relation can be expressed in terms of $v$,
\be\label{mpluseq}
\left. \frac{1}{f_+} \frac{\p M_+}{\p v} \right|_\Sigma = F(v) \, ,
\ee
with
\be\label{defF}
F(v) = \left. \frac{1}{f_-} \frac{\p M_-}{\p v} \right|_\Sigma \, . 
\ee
Eq.\ \eqref{mpluseq} constitutes a first order differential equation determining the dynamics of $M_+$ in terms of quantities given outside of the shell. Following Appendix \ref{app.A}, $M_+$ determines the curvature in the inside region. Thus, there is a close relation between $M_+$ and observable quantities in the inner region of the shell. For later convenience, it is useful to write the left-hand side of eq.\ \eqref{mpluseq}, stressing its dependence on $m_+$:
\be\label{defPoly}
\left. \frac{1}{f_+} \frac{\p M_+}{\p v} \right|_\Sigma = \frac{1}{P(m_+)} \, \dot{m}_+ \, . 
\ee
For the examples discussed in our work $P(m_+)$ is either linear or quadratic in $m_+$ with the coefficients of the polynomial depending on the model details.

Typically, the analysis of mass inflation considers a fixed background geometry of mass $m_0$, say, and adds a small perturbation to the mass function. The boundary condition at the event horizon is fixed through the Price tail behavior, so that \cite{Price,Price1}
\be\label{eq.price}
m_-(v) = m_0 - \frac{\beta}{(v/v_0)^p} \, 
\ee
in the region outside the shell. Here $\beta > 0$ is a quantity with the dimension of a mass and $v_0$ sets the initial value of $v$. Furthermore, the exponent governing the decay of the perturbation depends on its angular momentum and one has $p \ge 11$. 

The original studies of a fixed background geometry perturbed by an ingoing influx of energy appear to indicate that the Cauchy horizon is a generic surface of infinite blueshift \cite{Penrose1968}. Further investigations of gravitational wave perturbations near the Cauchy horizon confirmed the exponential divergence of the mass function and charged black holes with spherical symmetry, i.e., the Reissner-Nordstr\"om metric, have been shown to exhibit a curvature scalar singularity \cite{PI1990}. Furthermore Ori used an exact solution based on works of Poisson and Israel to show that the tidal forces at this singularity remain finite \cite{Ori:1991zz}. We stress that, strictly speaking, the Ori model building on the Price tail behavior \eqref{eq.price} is valid for asymptotically late times $v$ only. Therefore, conclusions drawn from the model in a regime where $v$ is small and the perturbation significant compared to the mass $m_0$ have to be interpreted with care and should be confirmed by an analysis of the full dynamics
\cite{Bonanno:1994qh}.

Recently it is been shown in \cite{Bonanno:2020fgp} that in fact the functional dependence of Misner-Sharp mass on $m_{+}$ determines the late-time behavior of these geometries. In particular, while a linear dependence leads to mass-inflation instability, a non-linear relation leading to a quadratic polynomial $P(m_+)$ in eq.\ \eqref{defPoly} can tame the exponential growth of $m_{+}$ in the late time regime. 

\section{Mass-inflation in the Hayward Geometry}
\label{sect.3}
Upon introducing our dynamical framework, we now specialize to the case of the Hayward geometry \cite{Hayward:2005gi}. The background geometry including the Hawking effect is introduced in Sect.\ \ref{sect.31} and a sketch of the conformal diagram related to the model is shown in Fig.\ \ref{fig.conformal}. The late-time dynamics of the model is discussed in Sect.\ \ref{sect.32} and we corroborate our findings through a numerical analysis in Sect.\ \ref{sect.33}. We then briefly comment on the strength of the curvature singularity in Sect.\ \ref{sect.34}.  
\subsection{The dynamical background geometry}
\label{sect.31}
For the Hayward geometry, the lapse function is given by
\be\label{eq.hay}
f(r) = 1 - \frac{2 M(r)}{r} \, , \quad M(r) = \frac{m r^3}{r^3 + 2 m l^2} . 
\ee 
Here $M$ is the Misner-Sharp mass and $m$ is the asymptotic mass of the configuration in Planck units where Newton's constant $G_N=1$. Furthermore, $l$ is a free parameter setting the scale where the modifications of the geometry set in. The Schwarzschild geometry is recovered for $l=0$ so that $M=m$ in this case.

The horizons of the geometry \eqref{eq.hay} appear as solutions of the horizon condition
\be
\label{condition}
2l^2m - 2mr^2 + r^3 = 0 \, . 
\ee
For sufficiently large values $m$, there is an (outer) event horizon at $r_+$ and an (inner) Cauchy horizon at $r_- < r_+$. Their explicit position as function of the parameters $m,l$ can be found by solving the cubic. The surface gravity at the horizons is defined as
\be\label{eq.surface}
\kappa_{\pm} \,=\, \pm \, \frac{1}{2} \left. \frac{\partial{f(r)}}{\partial{r}} \right|_{r=r_{\pm}} \, , 
\ee
where the signs are chosen such that $\kappa_\pm > 0$.
For the critical mass (indicated by the subscript \emph{cr})
\be\label{eq.crit}
m_{cr} = \frac{3\sqrt{3}}{4}l \, ,  
\ee
the two horizons coincide and are located at 
\be\label{def.rcrit}
r_{cr} = \sqrt{3}l \, .
\ee
In this case, one obtains a cold remnant.

\begin{figure}[t!]
	\centering
	\includegraphics[width = 0.45 \textwidth]{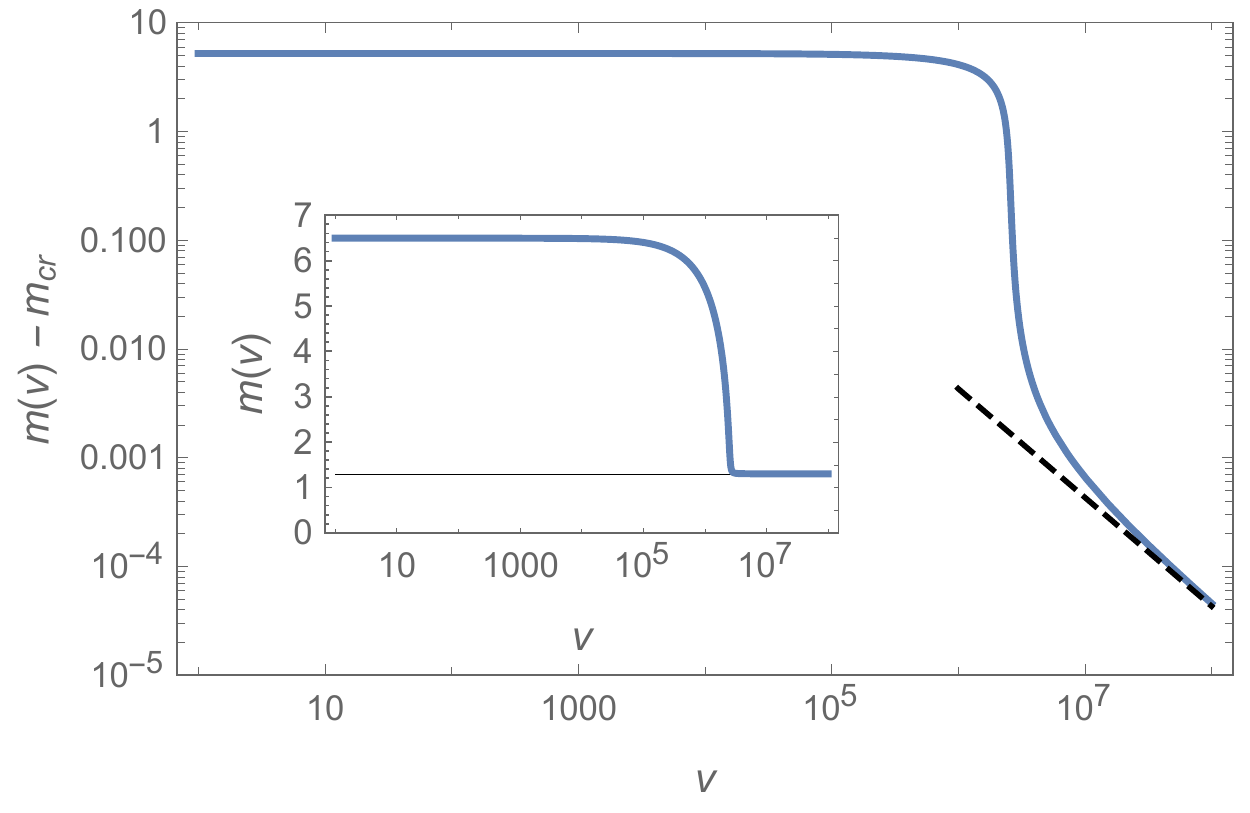} \\
	\includegraphics[width = 0.45 \textwidth]{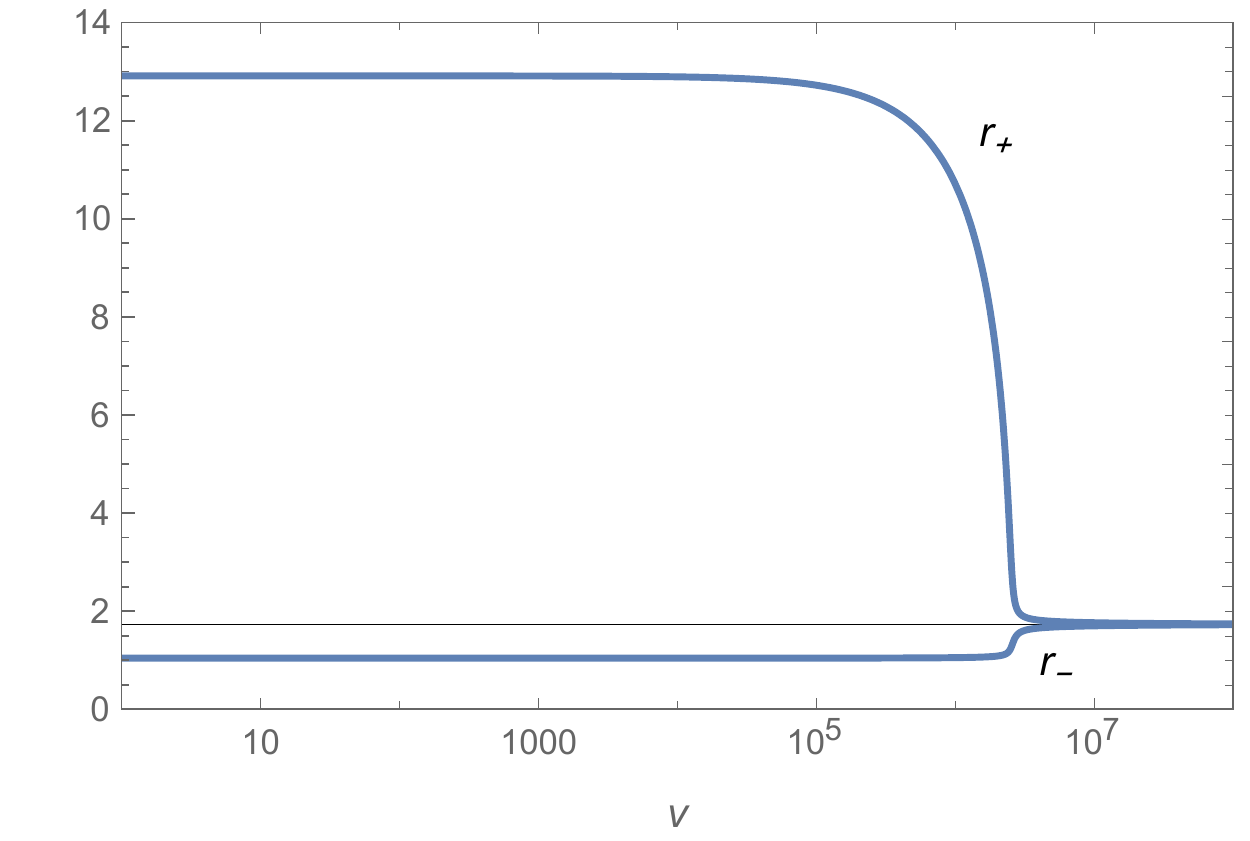}
	\caption{\label{Fig.1} Top diagram: illustration of $m(v)-m_{cr}$ resulting from numerically integrating eq.\ \eqref{eq:mass-loss}  with initial condition $m(1) = 5 m_{cr}$ and $l=1$ (blue line). The black dashed line marks the asymptotic solution \eqref{eq:mass-loss-asym}. For convenience, $m(v)$ approaching the critical mass $m_{cr}$ (black horizontal line) is depicted in the inset. Bottom diagram: radial position of the apparent event horizon $r_+(v)$ (top curve) and apparent Cauchy horizon $r_-(v)$ (bottom curve). Asymptotically, the horizons approach the critical value $r_{cr}$ depicted as the black horizontal line.}
\end{figure}

We now include the effect of the black hole evaporation through the emission of Hawking radiation. Following Hawking's seminal work \cite{Hawking:1975vcx}, the event horizon emits Hawking radiation with temperature
\be\label{HawkingTemp}
T = \frac{1}{4\pi} \left. \frac{\p f}{\p r} \right|_{r = r_+} \, . 
\ee
The power radiated from the black hole follows from Boltzmann's law
\be
P = \frac{\pi^2}{30} \, T^4 \, A \, , 
\ee
where $A \equiv 4 \pi r_+^2$ is the area of the event horizon. 
Promoting the temperature and horizon area to functions of $v$ and using the adiabatic approximation the mass-loss of the geometry can be computed from
\be\label{eq:mass-loss}
\frac{\p m(v)}{\p v} = - \frac{\pi^2}{30} \, T(v)^4 \, A(v) \,, 
\ee
where the area of the event horizon is approximated dynamically from the location of the 
apparent horizon in (\ref{condition}). Eq.\ 
\eqref{eq:mass-loss} turns into a closed equation determining the $v$-dependence of $m$. The late time behavior can be determined analytically,
\be\label{eq:mass-loss-asym}
m(v) \simeq m_{cr} + \frac{10935 \, l^4 \, \pi}{8 v} + \cO(v^{-3/2}) \, .
\ee
Here we have fixed the integration constant such that $m(v)$ approaches $m_{cr}$ for asymptotically large values $v$. The full solution of \eqref{eq:mass-loss} can be obtained by numerical integration and is shown in Fig.\ \ref{Fig.1}. 

At this point, it is instructive to compare eqs.\ \eqref{eq.price} and \eqref{eq:mass-loss-asym}. This reveals that perturbations following Price's law decay \emph{significantly faster} than the power-law corrections to the critical mass. This suggests that it is actually the Hawking effect which dominates the asymptotic behavior.
\subsection{Mass-inflation: identifying the late-time attractor}
\label{sect.32}
We are now in a position to evaluate the late-time dynamics of the Ori-model for the background given by the (time-dependent) Hayward geometry. This can be conveniently done by analyzing the equations of motion using the Frobenius method. It then turns out that the resulting scaling behaviors are crucially modified by cancellations of leading and subleading terms obtained at earlier stages in the expansion. In order to obtain the full information one therefore has to compute the leading as well as several subleading terms controlling the late-time dynamics of the mass-function $m_-(v)$ and $R(v)$. In order to avoid giving bulky formulas, we are displaying the leading terms in the expansion only and indicate the fact that we are dealing with asymptotic expressions by the $\simeq$-symbol. The full expressions and additional technical comments underlying the analysis can be found in Appendix \ref{App.B}.

We start with the mass-function in the region outside the shell, $m_-(v)$, adopting the late-time dynamics given in eq.\ \eqref{eq:mass-loss-asym}. Substituting a Frobenius ansatz shows that the asymptotic time-dependence is given by
\be\label{eq:mminushay}
m_-(v) \simeq m_{cr} + \frac{480 \pi \, m_{cr}^4}{v} \ . 
\ee
The time-dependent mass function also contains the contribution associated with the shell. The Price law behavior is subleading to the time-dependence induced by the Hawking effect though. Hence, this contribution is not visible in the asymptotic analysis.

In the next step, we specialize \eqref{Rpos} to the Hayward geometry,
\begin{equation}
\dot{R}(v)=\frac{1}{2}-\frac{m_{-} R^{2}}{R^{3}+2 l^{2} m_{-}} \, , 
\end{equation}
and determine the asymptotics of the position of the shell. At this point, it is instructive to first look at the right-hand side for $v$ fixed. The function $f_-$ vanishes at the ($v$-dependent) apparent event and Cauchy horizons. Thus the shell cannot cross these values. For the final configurations both $r_\pm(v)$ approach $r_{cr}$. Thus, we expect that the asymptotic position of the shell is given by $r_{cr}$. The leading corrections describing the approach of the shell can be found by making a Frobenius ansatz,
\be
R(v) = r_{cr} + \frac{1}{v^s} \sum_{n=0} r_n \, v^{-n} \, , 
\ee
where $s$ and $r_n$ are numerical coefficients. Substituting this expansion into eq.\ \eqref{Rpos} and demanding the existence of a non-trivial solution fixes $s=1/2$. This results in a quadratic equation for the numerical coefficient $r_0$ which has the solutions
$r_0 = \pm 3 \sqrt{15 \pi } \, r_{cr}^{5/2}$. Thus the late-time asymptotics of $R(v)$ is
\be\label{Rvasym}
R(v) \simeq r_{cr} \pm 3 \, r_{cr} \, \left( \frac{15 \pi \, r_{cr}^3}{v} \right)^{1/2}\, . 
\ee
The sign-ambiguity indicates that the shell can approach $r_{cr}$ either from above $r > r_{cr}$ or below $r < r_{cr}$. Both cases can be physically viable a priori.

Subsequently, we substitute eqs.\ \eqref{eq:mminushay} and \eqref{Rvasym} into \eqref{mpluseq}. This yields
\begin{equation} \label{att1}
\frac{\dot{m}_{+}}{\left(R^{3}+2 l^{2} m_{+}\right)\left(R^{3}-2 m_{+}\left(R^{2}-l^{2}\right)\right)}=F(v)
\end{equation}
with
\begin{equation} \label{rhs}
F(v) \equiv \frac{\dot{m}_{-}}{\left(R^{3}+2 l^{2} m_{-}\right)\left(R^{3}-2 m_{-}\left(R^{2}-l^{2}\right)\right)} \, .
\end{equation}
This equation exhibits two values for $m_+$ where $\dot{m}_+$ vanishes. Keeping $v$ fixed and finite, these values are given by the roots of the quadratic polynomial
\begin{equation} \label{att2}
P(m_+) = \left(R^{3}+2 l^{2} m_{+}\right)\left(R^{3}-2 m_{+}\left(R^{2}-l^{2}\right)\right) \, . 
\end{equation}

Substituting the asymptotic expansions of $R(v)$ and $m_-(v)$ into eq.\ \eqref{rhs} gives the following asymptotics for $F(v)$:
\be\label{Fasymp}
F(v) \simeq \pm 4 m_{cr} \, \left( \frac{5 \pi m_{cr}}{v} \right)^{1/2}. 
\ee
Hence, the right-hand side of eq.\ \eqref{att1} vanishes as $v \rightarrow \infty$. As a consequence $m_+(v) \simeq {\rm const}$ is an asymptotic solution. In order to fix this constant,  we then solve the differential equation
\be
\frac{27 \, l^2 \, \dot{m}_+}{27 l^2-6 \sqrt{3} l m_+-8 m_+^2} = \pm 4 m_{cr} \, \left( \frac{5 \pi m_{cr}}{v} \right)^{1/2}
\ee
which results from substituting the asymptotic expansions  \eqref{eq:mminushay}, \eqref{Rvasym} and \eqref{Fasymp} into \eqref{mpluseq} and retaining the terms leading in $1/v$ only. Taking into account the sign-ambiguity in \eqref{Rvasym}, the resulting asymptotic solutions for $m_+$ are
\be
m_+^{\rm asym} = -
\frac{m_{cr}}{2} \left(1 \mp 3  \tanh \left(6 \sqrt{5 \pi \,  
m_{cr} \,v}+c\right)\right) \, , 
\ee
where $c$ is an integration constant. This constant drops out in the limit where $v \rightarrow \infty$. Thus, for a shell approaching $r_{cr}$ from above 
\be\label{mplusin}
\lim_{v \rightarrow \infty} m_+(v) = m_{cr} \, , 
\ee
while a shell approaching $r_{cr}$ from below leads to the attractor
\be\label{mplusin2}
\lim_{v \rightarrow \infty} m_+(v) = - 2 m_{cr} \, .
\ee
These correspond to the two roots of \eqref{att2} when evaluated at asymptotically late times. In particular, eq.\ \eqref{mplusin2} generalizes the attractor taming mass-inflation in the non-dynamical setting \cite{Bonanno:2020fgp}, to the case of an evaporating black hole. The subleading correction to \eqref{mplusin2} is found by the Frobenius analysis described in App.\ \ref{App.B} and reads
\be\label{HWmpasym}
m_+(v) \simeq -2 m_{cr} + 48 \sqrt{\frac{ 5 \pi m_{cr}^5}{v}} \, . 
\ee

We proceed with analyzing the asymptotics of the physical quantities including the curvature scalars computed in Appendix \ref{app.A}. We distinguish the cases where the shell approaches $r_{cr}$ from above (upper signs) and below (lower signs). We start by discussing the upper sign. In this case the Misner-Sharp mass remains finite
\be\label{MSfinite}
\left. M_+(r,v) \right|_\Sigma \simeq \frac{2 m_{cr}}{3} \, . 
\ee
Consequently, the curvature invariants remain finite as well and evaluate to
\be
\begin{split}
\left. C^2 \right|_{\Sigma} \simeq & \, \frac{2160 \pi }{m_{cr} v} \, , \quad \left. K \right|_{\Sigma} \simeq \, \frac{81}{32 \, m_{cr}^4} \, . 
\end{split}
\ee
The $v$-independent terms in this expansion agree with the curvature scalars evaluated at the critical radius of the critical configuration. Hence, any perturbation which falls onto $r_{cr}$ \emph{from above} does not have any destabilizing effects on the geometry.

For the lower sign, the asymptotic behavior drastically differs. Substituting the asymptotic expansions into the general expression of the Misner-Sharp mass and evaluating at the position of $\Sigma$, we have
\be\label{MSdiv}
\left. M_+(r,v) \right|_\Sigma \simeq - v \, . 
\ee
While $M_+(r,v)$ is generically finite, it diverges in the limit $v \rightarrow \infty$ when evaluated at $r = r_{cr}$ due to a cancellation of the leading terms in the denominator. This divergence is much milder than the one encountered in the standard mass-inflation effect where $M_+$ grows exponentially in $v$. 

The divergence \eqref{MSdiv} also propagates into the scalar curvature invariants. In particular
\be\label{HW.curvature}
\begin{split}
\left. C^2 \right|_{\Sigma} =  \, \frac{19683 \, v^6}{4096 \,  m_{cr}^{10}}\, , \qquad 
\left. K \right|_{\Sigma} = & \, \frac{59049 \, v^6}{4096 \,  m_{cr}^{10}} \, . 
\end{split}
\ee
Thus, again, the divergence is power-law and not exponential. Remarkably, the results \eqref{MSdiv} and \eqref{HW.curvature} are \emph{universal} in the sense that they are independent of any free parameter. Hence any solution entering into this late-time scaling regime must follow this attractor behavior. 
\subsection{Mass-inflation: full numerical treatment}
\label{sect.33}
\begin{figure}
	\includegraphics[width=0.45\textwidth]{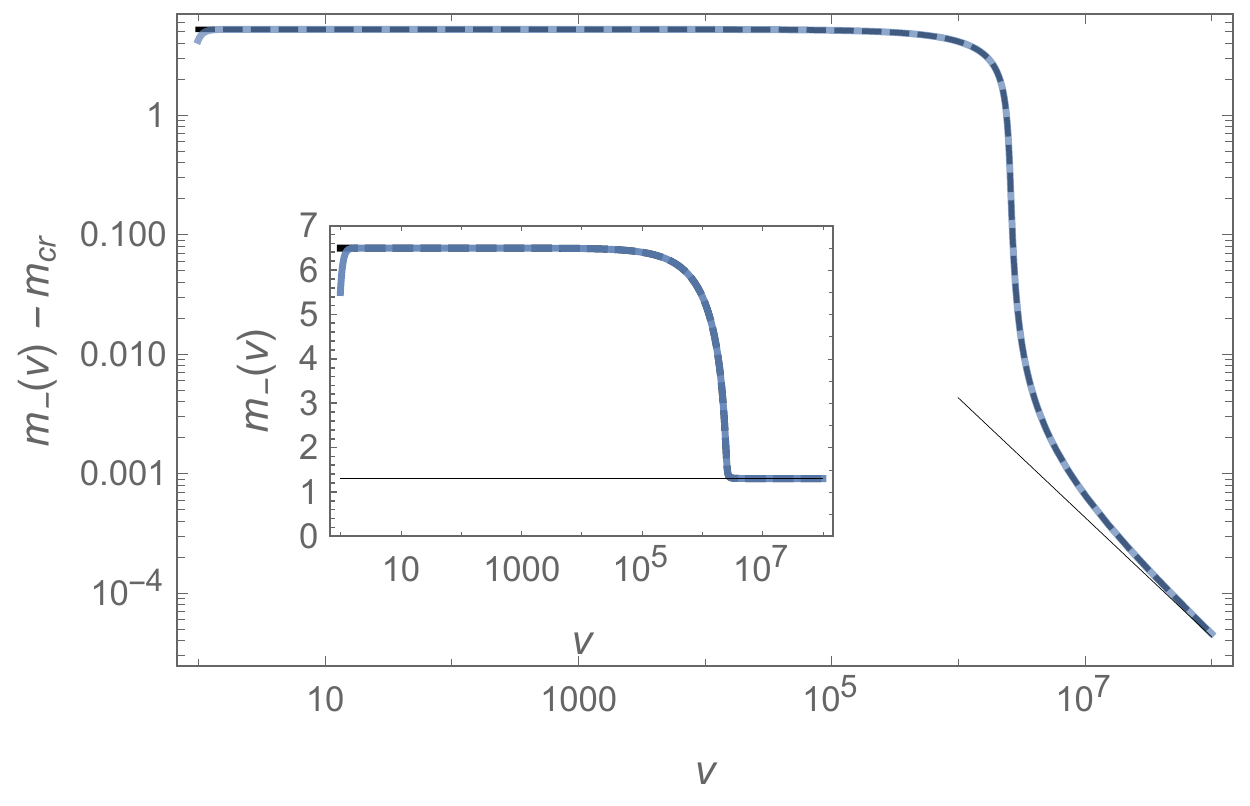}
	\caption{Illustration of the background geometry $m_-(v)-m_{cr}$ resulting from \eqref{mminusback} for $\beta = 1$ (blue curve) and $\beta = 0$ (dashed) for $l=1$ in a log-log-plot. The solid line gives the asymptotic behavior \eqref{eq:mass-loss-asym}. Inset: $m_-(v)$ for $\beta = 1$ (blue curve) and $\beta = 0$ (dashed) with the solid line showing $m_{cr}$. The perturbation due to the Price-tail corrections vanishes at very short time-scales so that the lines agree for $v \gtrsim 10$. \label{mminusbackground}}
\end{figure}

Following up on the asymptotic analysis of the previous section, we proceed with a numerical investigation of the dynamics. We start by constructing $m_-(v)$ in the outer sector of the shell by superimposing the mass-loss due to Hawking radiation and the perturbation due to the shell
\be\label{mminusback}
m_-(v) = m_{\rm Hawking}(v) - \frac{\beta}{v^p} \, .
\ee
Here $m_{\rm Hawking}(v)$ is obtained from solving eq.\ \eqref{eq:mass-loss} numerically with the boundary condition $\lim_{v \rightarrow \infty} m_{\rm Hawking}(v) = m_{cr}$ according to eq.\ \eqref{eq:mass-loss-asym}. For concreteness, we take $\beta = 1$, $l=1$, and $p=11$. The resulting function $m_-(v)$ is then depicted in Fig.\ \ref{mminusbackground}.
This shows that the Price tail perturbation is only operative at values $v$ much smaller than the time-scale set by the Hawking effect. This is in agreement with the result of the previous subsection that the late-time behavior of $m_-(v)$ is controlled by the Hawking effect only.

\begin{figure}
	\includegraphics[width=0.45\textwidth]{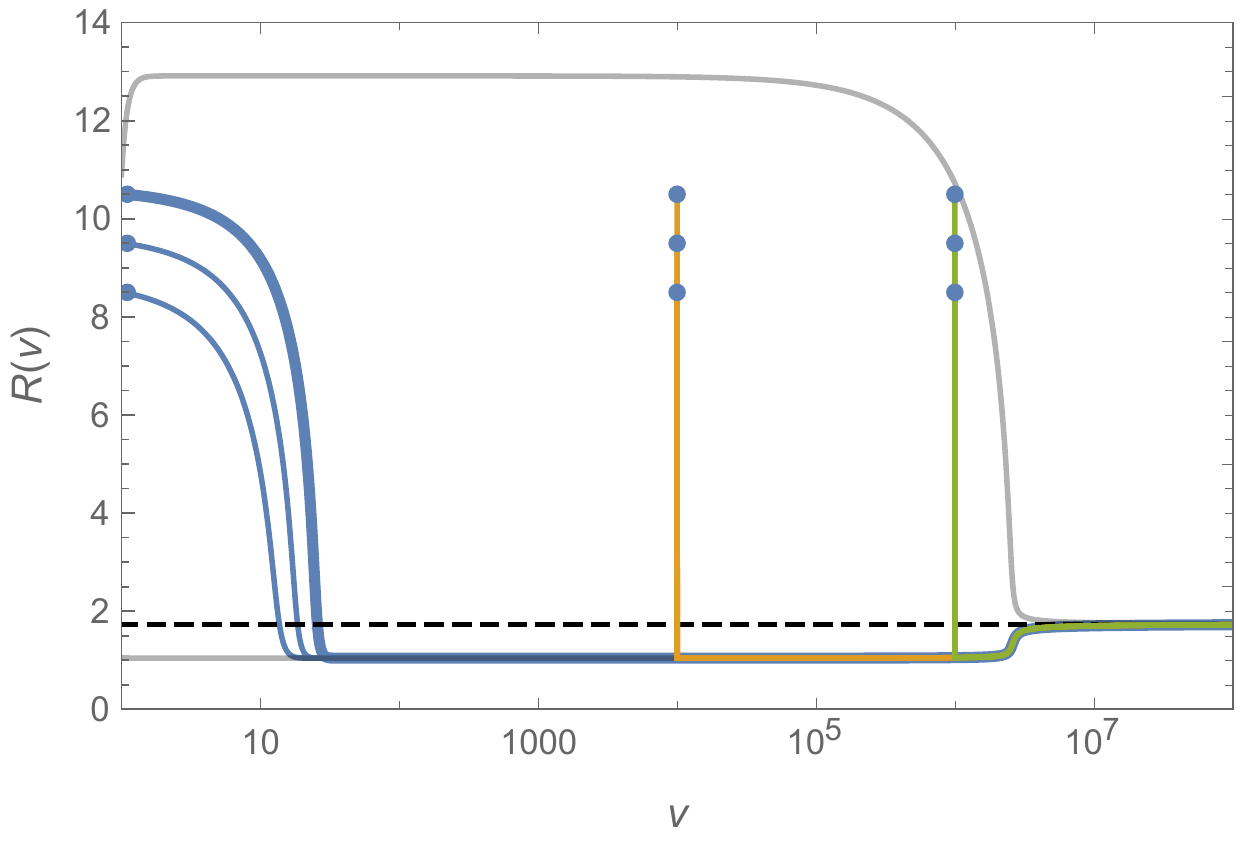}
	\caption{Typical trajectories $R(v)$ describing the motion of shells with initial conditions marked by the blue dots at $v=1$ (blue series), $v=10^4$ (orange) and $v=10^6$ (green) in a background where $m_{\rm Hawking}(1) = 5 m_{cr}$. The apparent event and Cauchy horizons are depicted as opaque gray lines while $r_{cr}$ is given by the dashed line. All shells settle at $r_{cr}$ at asymptotically late times. \label{shellposition}}
\end{figure}

Given $m_-(v)$, we solve eq.\ \eqref{Rpos} to trace the position of the shell in the background geometry. We consider a shell situated between the apparent event horizon and Cauchy horizon. Choosing the initial mass of the unperturbed configuration as $m_{\rm Hawking}(1) = 5 m_{cr}$ and imposing initial conditions at $v_{init} = \{1, 10^4, 10^6\}$, typical trajectories $R(v)$ are displayed in Fig.\ \ref{shellposition}. 
The solutions exhibit several phases which are universal in the sense that they are independent of the initial conditions. At early times, the shell falls towards smaller values of $r$, approaching the apparent Cauchy horizon of the time-dependent geometry. Subsequently, the solutions trail $r_-(v)$ until they settle on the late-time attractor \eqref{Rvasym}. Note that for late times we have $\dot{R}(v) > 0$ indicating that the late-time dynamics is actually governed by the minus-sign in eq.\ \eqref{Rvasym}.

Specifying \eqref{mpluseq} to the Hayward geometry then leads to the following equation determining the mass function $m_+(v)$ inside the shell
\be\label{mplushay}
\dot{m}_+=\frac{\left(2 l^2 m_+ +R^3\right) \left(R^3-2 \left(R^2-l^2\right) m_+ \right)}{\left(2 l^2 m_-+R^3\right) \left(R^3-2 \left(R^2-l^2\right) m_- \right)} \, \dot{m}_- \, . 
\ee
Substituting the solutions for $m_-(v)$ and the dynamics $R(v)$, allows to solve this equation numerically. For the latter, we adopt select the shell moving along the top (blue) curve in Fig.\ \ref{shellposition}.
%
\begin{figure}
	\includegraphics[width=0.45\textwidth]{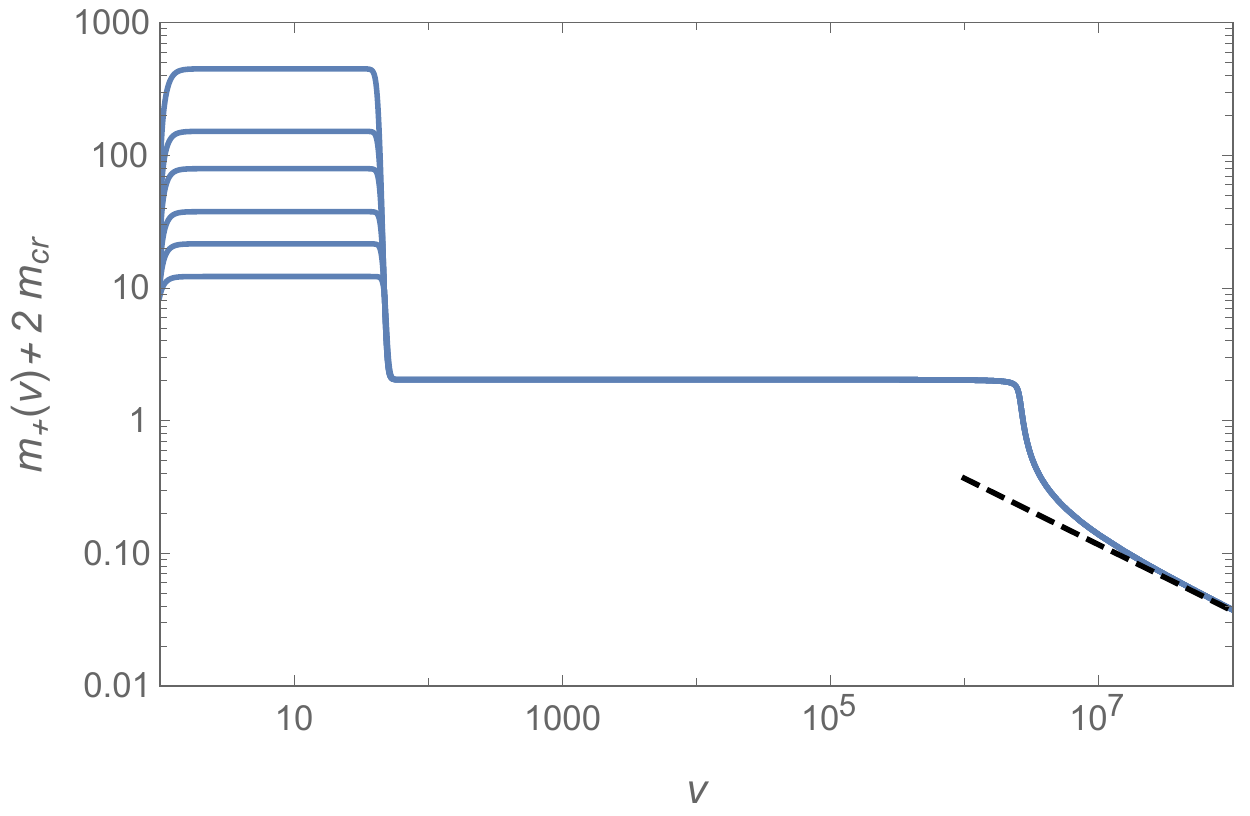}
	\caption{Time-evolution of $m_+(v)$ arising from solving \eqref{mplushay} in the background provided by \eqref{mminusback} and the shell moving along the thick blue curve depicted in Fig.\ \ref{shellposition}. Initial conditions are imposed at $v=1$ and chosen such that $m_+(1)$ is larger than the critical value required for the solutions reaching the late-time attractor  \eqref{mplusin2}. The dashed line indicates the asymptotic scaling behavior \eqref{HWmpasym}. \label{mplusdyn}}
\end{figure} 
Depending on the initial value for $m_{+}^{init}$, the numerical integration identifies two classes of solutions. Solutions starting with $m_{+}^{init} < m_+^{crit}$ terminate at finite $v$. Solutions where $m_{+}^{init} \ge m_+^{crit}$ exist for all values $v$ and follow the attractor \eqref{mplusin2} at late times. A series of solutions taken from the class $m_{+}^{init} \ge m_+^{crit}$  are illustrated in Fig.\ \ref{mplusdyn}. Notably, the information on the initial conditions is wiped out rather quickly as the shell settles on $r_-(v)$. As a result the dynamics becomes essentially universal for $v \ge 100$. For late times the solutions settle on the attractor \eqref{HWmpasym}.

\begin{figure}
	\includegraphics[width=0.45\textwidth]{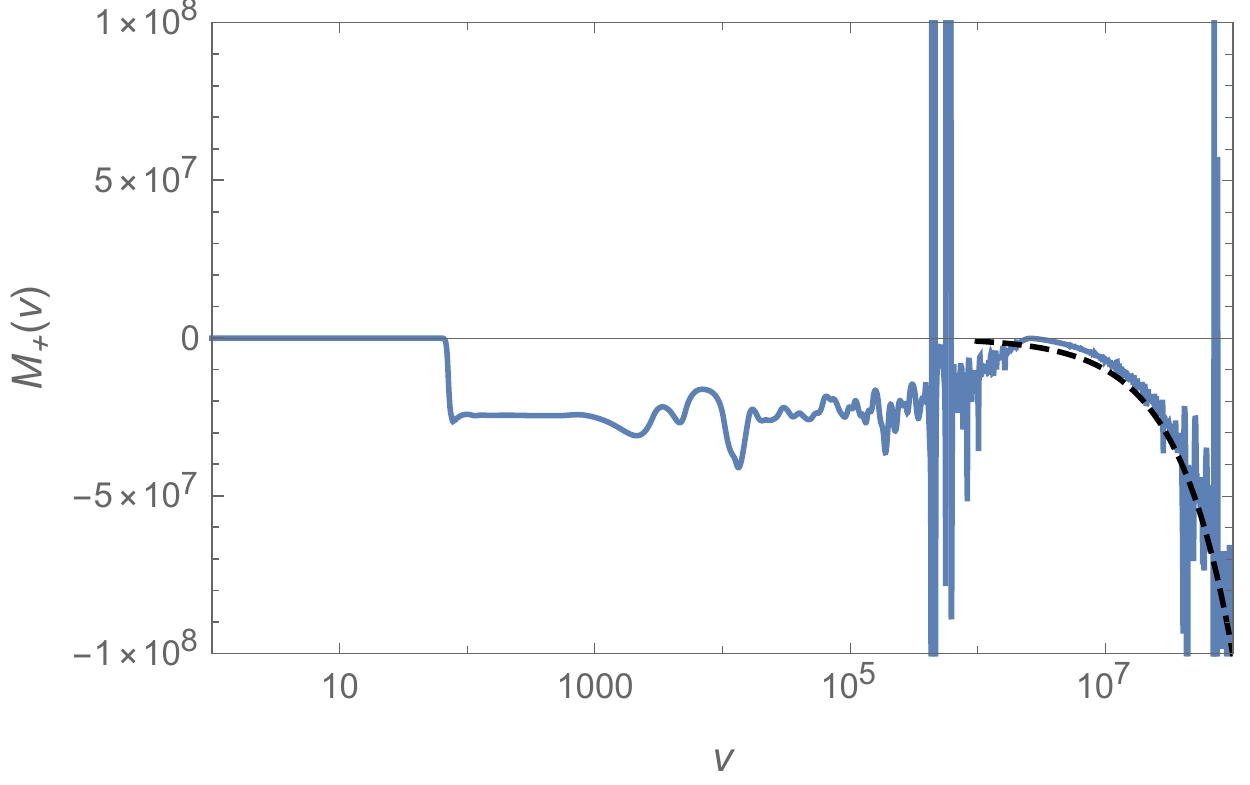}
	\caption{Time-evolution of the Misner-Sharp mass evaluated in the inner sector of the shell. The rapid increase in the curvature is triggered by the shell aligning with the dynamical Cauchy horizon. The attractor behavior taming $M_+(v)|_\Sigma$  sets in when the background reaches the final stage of black hole evaporation and is preceded by a rapid decrease of $M_+(v)|_\Sigma$. The late-time attractor \eqref{MSdiv} is added as the dashed line. The wiggles exhibited by $M_+(v)$ should be interpreted as numerical artifacts. Their occurrence can be traced by to the denominator structure of $M_+(v)$ where the dynamics is essentially governed by the cancellation of two numerical quantities of order unity to very high precision. Hence $M_+(v)$ is particularly sensitive to small errors in the numerical solutions of the differential equations. While this shows up in local wiggles, this does not affect the overall dynamics though. \label{FigMplus}}
\end{figure}

\begin{figure}
	\includegraphics[width=0.45\textwidth]{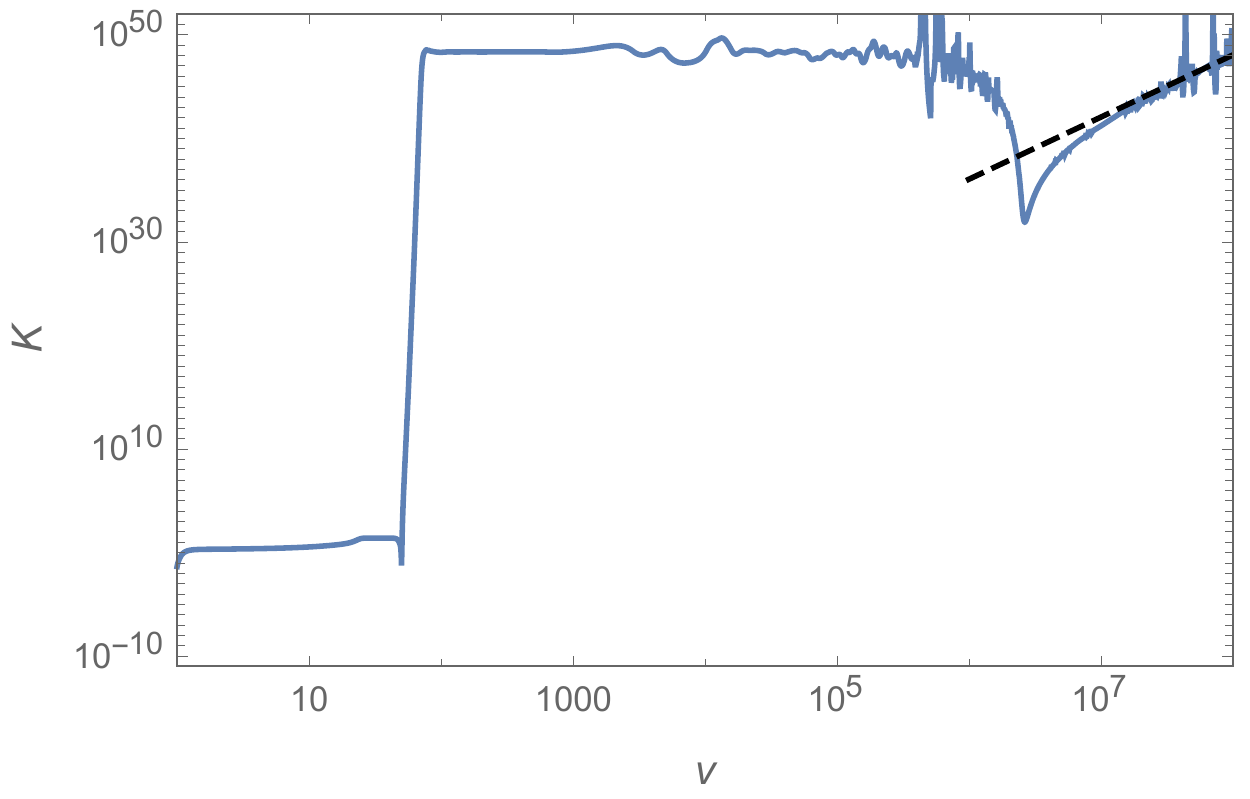}
	\includegraphics[width=0.45\textwidth]{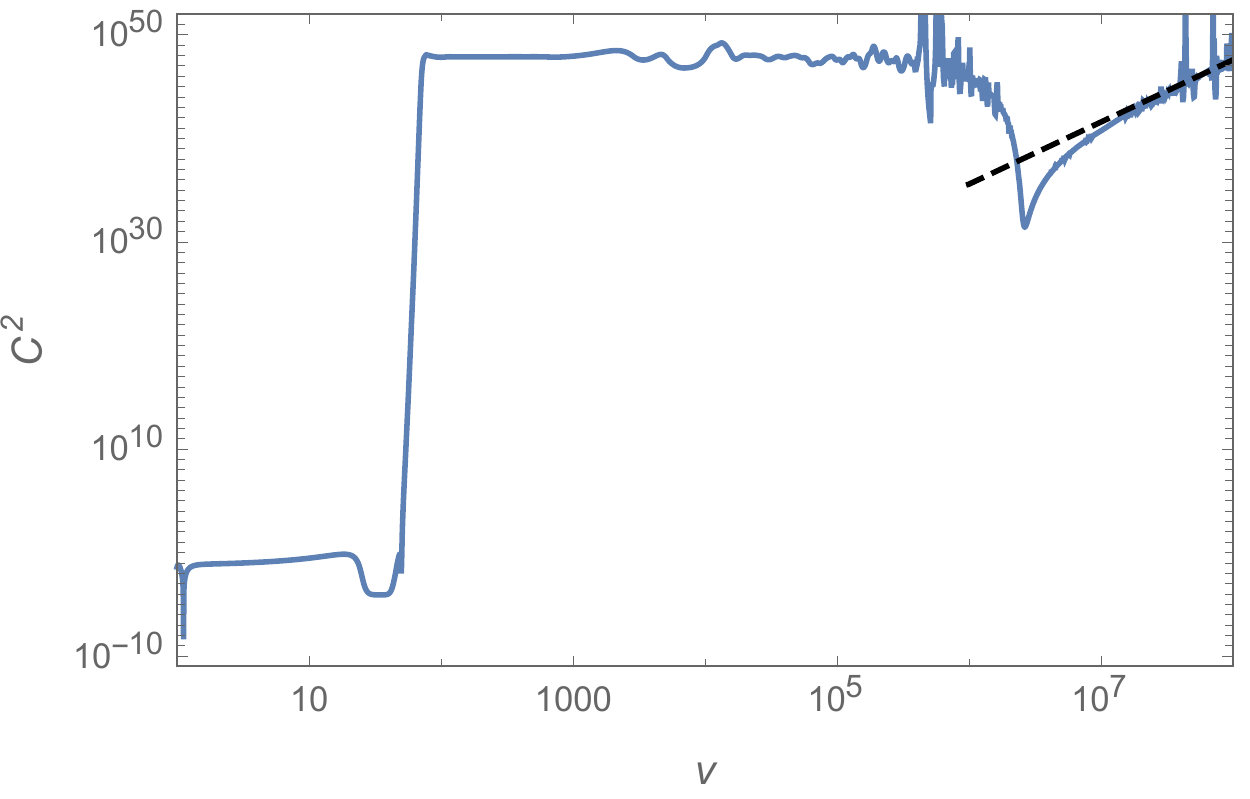}
	\caption{Time-evolution of the curvature scalars \eqref{Kev} and \eqref{C2ev} evaluated in the inner sector of the shell. The rapid increase in the curvature is triggered by the shell aligning with the dynamical Cauchy horizon. The power-law scaling \eqref{HW.curvature} has been added as a dashed line. Again the wiggles on the curve should be understood as numerical artifacts resulting from precise cancellations of quantities computed numerically. \label{FigCurvatureHay}}
\end{figure}

Given the explicit expression for $m_+(v)$, it is now straightforward to determine the resulting Misner-Sharp mass $M_+(v)|_{\Sigma}$ and the explicit form of the curvature scalars \eqref{Kev} and \eqref{C2ev}. This data is shown in Figs.\ \ref{FigMplus} and \ref{FigCurvatureHay}, respectively.
The Misner-Sharp mass undergoes a transition from small to rather large values as the shell impacts onto $r_-(v)$. This can be traced back to the denominator in \eqref{eq.hay} becoming small (but not zero though). This behavior is transmitted into the curvature scalars which undergo a rapid increase during this transition before remaining at large constant values.
This feature reflects the exponential grows of the Misner-Sharp mass reported in \cite{Carballo-Rubio:2021bpr,DiFilippo:2022qkl}. It precedes the attractor behavior for the curvature reported in the previous section.

At this point we stress that strictly speaking initial conditions for the dynamics should only be imposed at late time where the Ori model is capable of providing an accurate description of dynamics. The global analysis in this section clearly reveals that extrapolating the Ori model to early times may lead to misleading conclusions about the late-time behavior. This is particularly striking for solutions which hit a singularity before actually reaching the late-time attractor of the model.

\subsection{Strength of the curvature singularity}
\label{sect.34}
It is interesting to discuss the implications of our findings with regard to possible extensions
of the geometry beyond the Cauchy horizon. In the standard  Poisson-Israel model \cite{PI1990}, as 
first realized by \cite{Ori:1991zz} and further discussed by \cite{Burko:1999zv}, 
the Cauchy singularity does not meet the necessary conditions to be strong
in the Tipler sense \cite{Tipler:1977zza}. In particular the tidal distortion experienced by an observer crossing the singularity is finite. In fact it is not difficult to find
a coordinate transformation that renders the metric regular in the future sector
of the shell \cite{Bonanno:1994qh}. However, the singularity is still strong in the 
Kr\'olak sense because the expansion of the null congruence is (negatively) diverging at the Cauchy horizon \cite{Burko:1999zv}. 

In the case of regular black holes the exponential divergence of the classical
Poisson-Israel model is modified into a power-law divergence of the type
$\sim v^{6p}$, with $v\rightarrow\infty$ where $p\ge 11$,  and the singularity
is Kr\'olak-weak \cite{Bonanno:2020fgp}. Eq.\ \eqref{HW.curvature} confirms 
this picture once the effect of the Hawking flow, corresponding to $p=1$ in the notation \eqref{eq.price}, is taken into account. This supports the idea  that the strength of the inner horizon singularity in regular black holes
is much  milder than in the standard case. Moreover, it is completely
fixed by the properties of the Hawking radiation at late advanced times. 
On the other hand, because our geometry is very close to the extremality, it is not difficult to
show that, at variance with the non-extremal case,
this singularity is never reached in a finite amount of proper time
because $\tau \propto v$ as discussed in Appendix \ref{app.C}. 
\section{Mass-inflation for the Reissner-Nordstr{\"o}m geometry revisited}
\label{sect.4}
An important insight obtained from the previous section is that the late-time behavior of $m_-(v)$ and $R(v)$ is controlled by the Hawking effect. In particular, the dynamics induced by the time-dependent background changes the late-time behavior from $\dot{R}(v) < 0$ to $\dot{R}(v) > 0$. This qualitative change warrants revisiting the mass-inflation effect in the (classical) Reissner-Nordstr{\"o}m geometry, where it has been established initially using a static background \cite{PI1990,Ori:1991zz}. The  analysis follows the steps of the previous section. As the main result, we demonstrate that the mass-inflation is absent once the dynamics of the background is introduced.
%
\subsection{Geometry and late-time attractors}
\label{sect.41}
The line-element of the Reissner-Nordstr{\"o}m geometry is again of the form \eqref{sphericalsymline} with the lapse function
\be\label{lineRN}
f(r) = 1-\frac{2m}{r} + \frac{Q^2}{r^2} \, , \quad M = m - \frac{Q^2}{2 r} \, .
\ee
Here $Q$ is the electric charge of the configuration. Similarly to the Hayward geometry, the Reissner-Nordstr{\"o}m geometry possesses an event horizon and a Cauchy horizon situated at
\be\label{rs.horizons}
\begin{split}
r_\pm = m \pm \sqrt{m^2-Q^2} \, . 
\end{split}
\ee
For $m^2 = Q^2$ the two horizons coincide and one obtains an extremal black hole with
\be\label{rsextreme}
m_{cr} = Q \, , \quad r_{cr} = Q \, . 
\ee
For $m^2 > Q^2$ the event horizon again emits Hawking radiation with temperature given by \eqref{HawkingTemp}. This leads to a mass-loss described by \eqref{eq:mass-loss}. 

The dynamics of the shell and geometry in its interior sector is again captured by eqs.\ \eqref{Rpos} and \eqref{mpluseq}. Adapting these to the geometry \eqref{lineRN}, one arrives at the following system of coupled differential equations
\be\label{shelldynamicsRN}
\begin{split}
\dot{R} = & \, \frac{R^2+Q^2 -2 R \, m_-}{2R^2} \, , \\
 \dot{m}_+ = & \, \frac{R^2+Q^2 -2 R\,  m_+ }{R^2} \, F(v) \, , 
\end{split}
\ee
where
\be\label{FRN}
F(v) = \frac{R^2 \, \dot{m}_-}{R^2+Q^2 -2 R \, m_- } \, . 
\ee
Note that we have suppressed the $v$-dependence of $m_-$ arising from the Hawking effect and the Price tail perturbation.

Determining the asymptotics of the resulting solutions based on the Frobenius method is then rather straightforward. The mass function approaches $m_{cr}$ with the leading corrections falling off proportional to $1/v$,
\be
m_-(v) \simeq m_{cr} + \frac{30 \pi \, m_{cr}^4}{v} + O(v^{-3/2}) \, . 
\ee
The shell  approaches $r_{cr}$ according to
\be
R(v) \simeq r_{cr} -  \frac{2 \sqrt{15 \pi} \,  r_{cr}^{5/2}}{\sqrt{v}} + O(v^{-1}) \, . 
\ee
The last relation already anticipates that the shell crosses $r_{cr}$ and subsequently approaches the critical radius from $R(v) < r_{cr}$. Based on these relations, one obtains that
\be
F(v) \simeq - \sqrt{\frac{15 \pi \, m_{cr}^3}{v}}
+O(v^{-1}) \, , 
\ee
i.e., $F(v)$ vanishes asymptotically. This has profound consequences for the mass function in the inner sector of the shell. Consistency of \eqref{shelldynamicsRN} requires that
\be\label{RN.mass}
m_+(v) \simeq m_{cr} +\frac{30 \pi  m_{cr}^4}{v} \, , 
\ee
which goes hand-in-hand with $(Q^2 + R^2 - 2 R m_+) \simeq \cO(v^{-3/2})$. 

The crucial difference to the Hayward model then comes with the explicit form of the scalar curvature invariants tabulated in Table \ref{Tab.3}. For the Reissner-Nordstr{\"o}m case the denominators appearing in $K|_\Sigma$ and $C^2|_\Sigma$ remain finite. As a result the asymptotic limit of both curvature scalars \emph{remains finite} as $v \rightarrow \infty$:
\be\label{RN.curvature}
\begin{split}
K|_\Sigma \simeq & \, \frac{8}{m_{cr}^4} + \cO(v^{-1/2}) \, , \\
C^2|_\Sigma \simeq & \, \frac{2880 \pi }{m_{cr}  \, v} + \cO(v^{-3/2}) \, .
\end{split}
\ee
This is in striking difference with the standard mass-inflation effect where the curvature grows exponentially in $v$.

We now solve the system of equations \eqref{shelldynamicsRN} numerically, using a superposition of the Hawking radiation and Price tail perturbation for $m_-(v)$. We first consider a shell with initial conditions imposed at $v=1$, even though this is out of the validity-range of the Ori model. The motion of this shell is quantitatively identical to the Hayward case shown in Fig.\ \ref{shellposition}. The shell starts from its initial point, quickly falls towards the apparent Cauchy horizon, and eventually traces $r_-(v)$ into the asymptotic scaling regime. 

\begin{figure*}
	\includegraphics[width=0.45\textwidth]{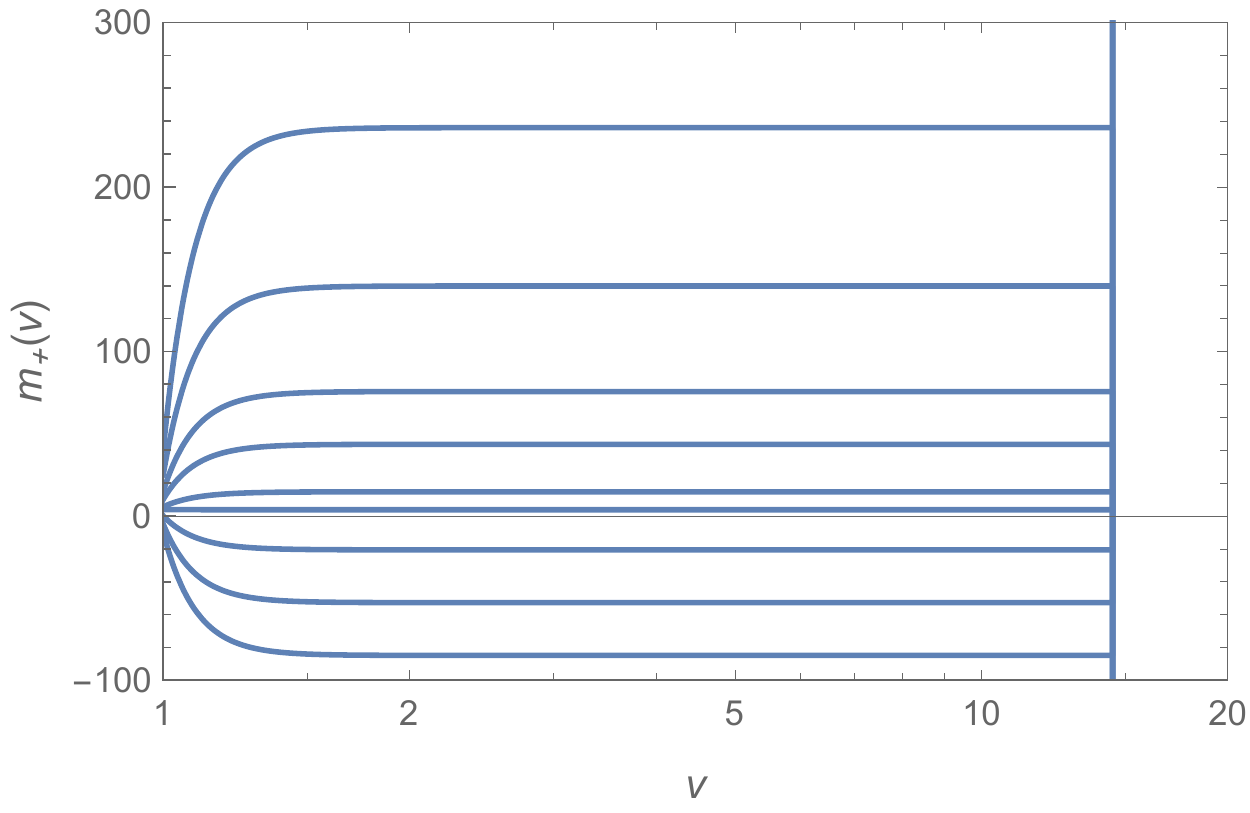} \, \,
	\includegraphics[width=0.45\textwidth]{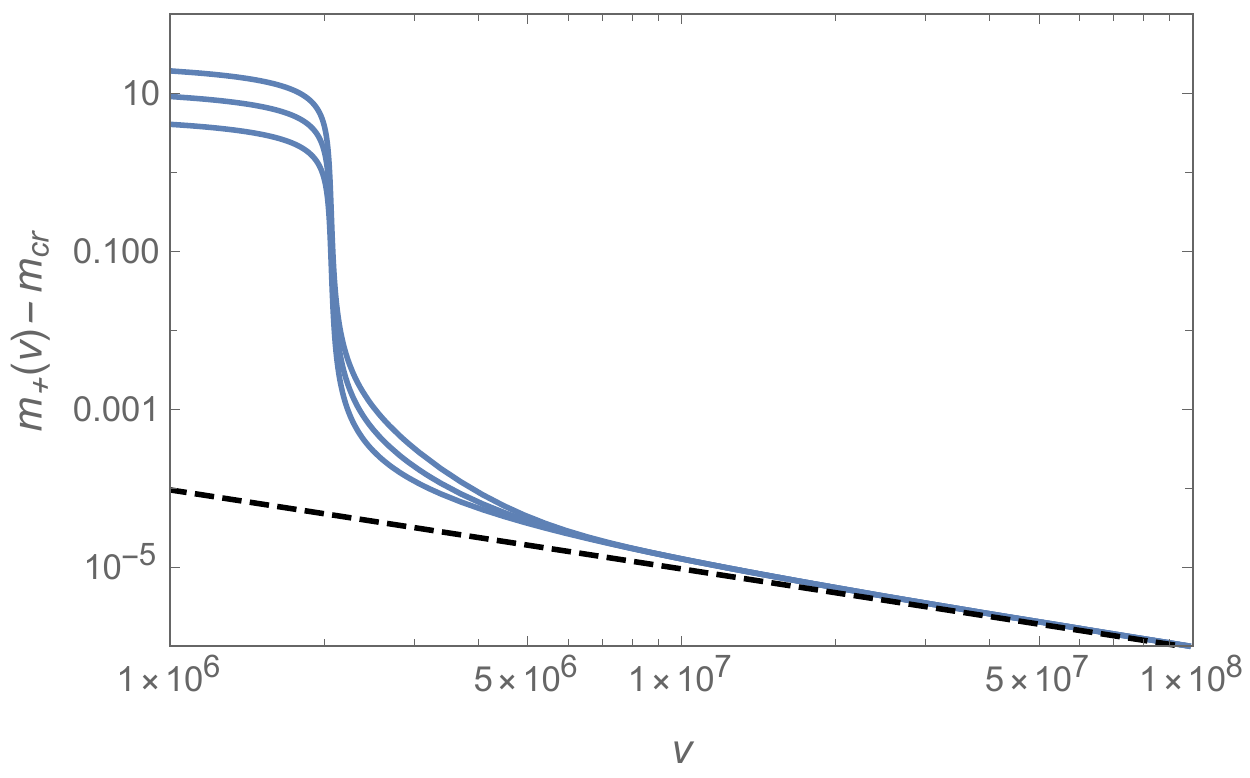} \\
	\includegraphics[width=0.45\textwidth]{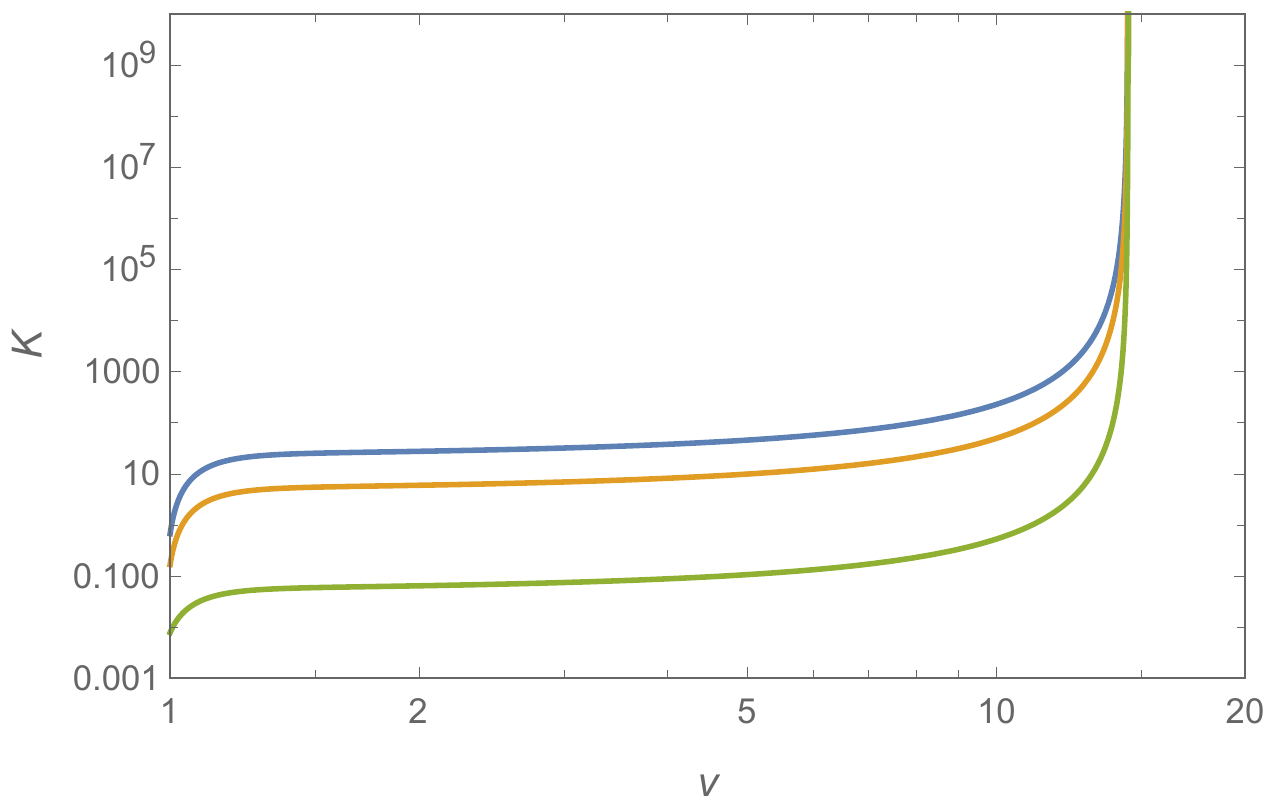} \, \,
		\includegraphics[width=0.45\textwidth]{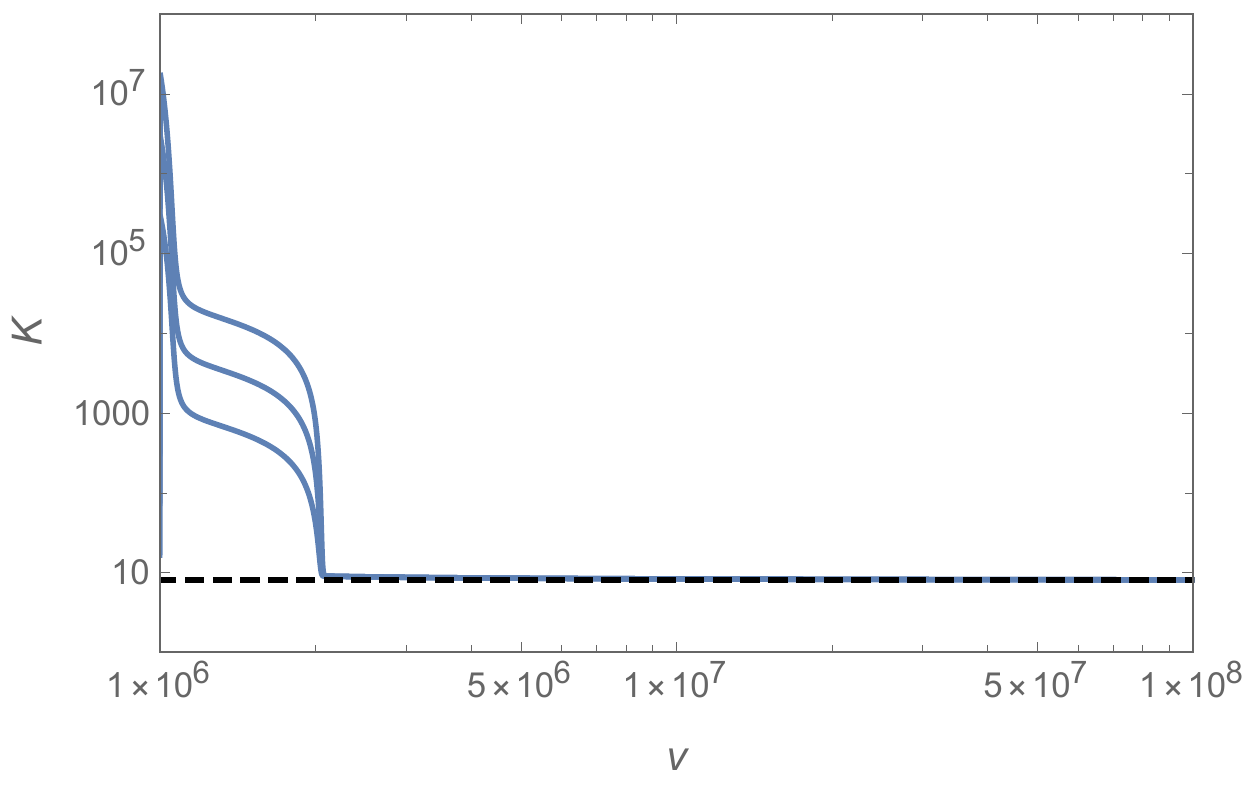} \\
		\includegraphics[width=0.45\textwidth]{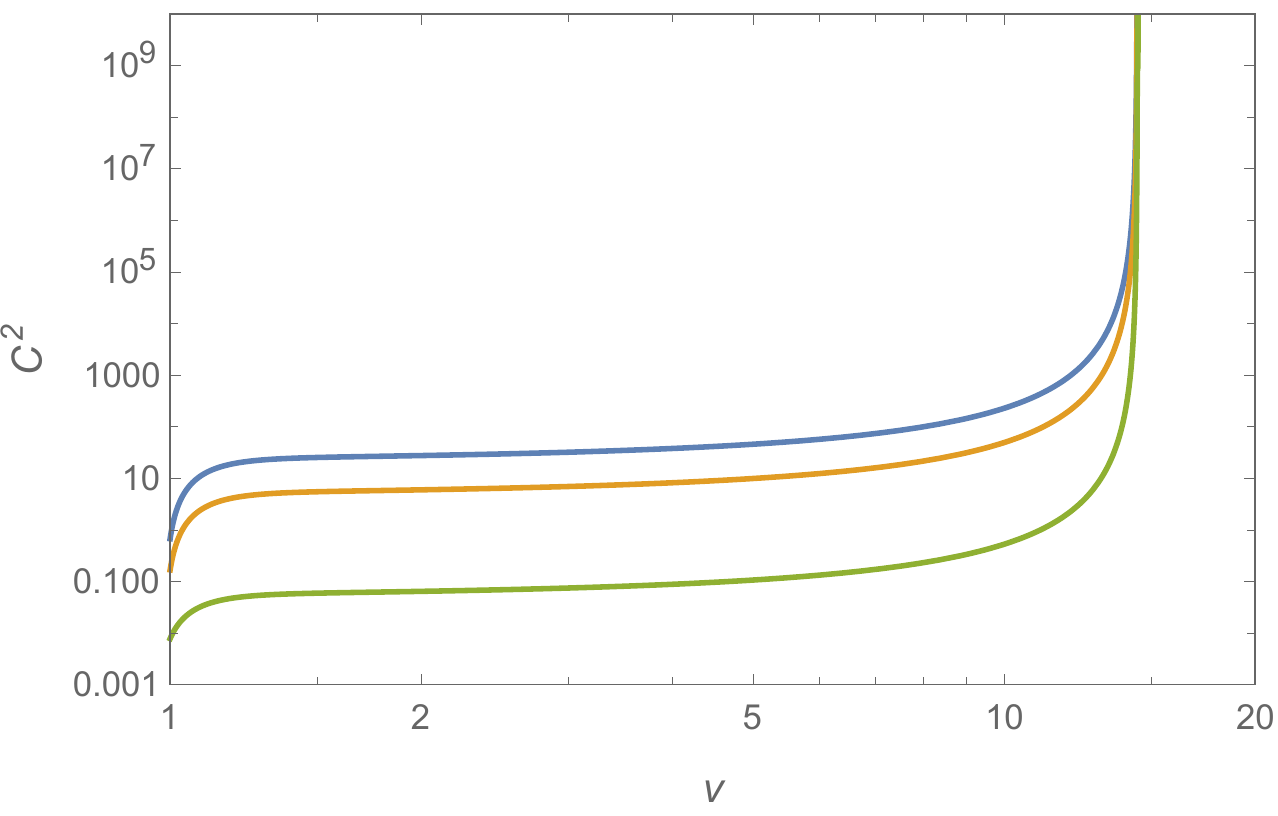} \, \,
		\includegraphics[width=0.45\textwidth]{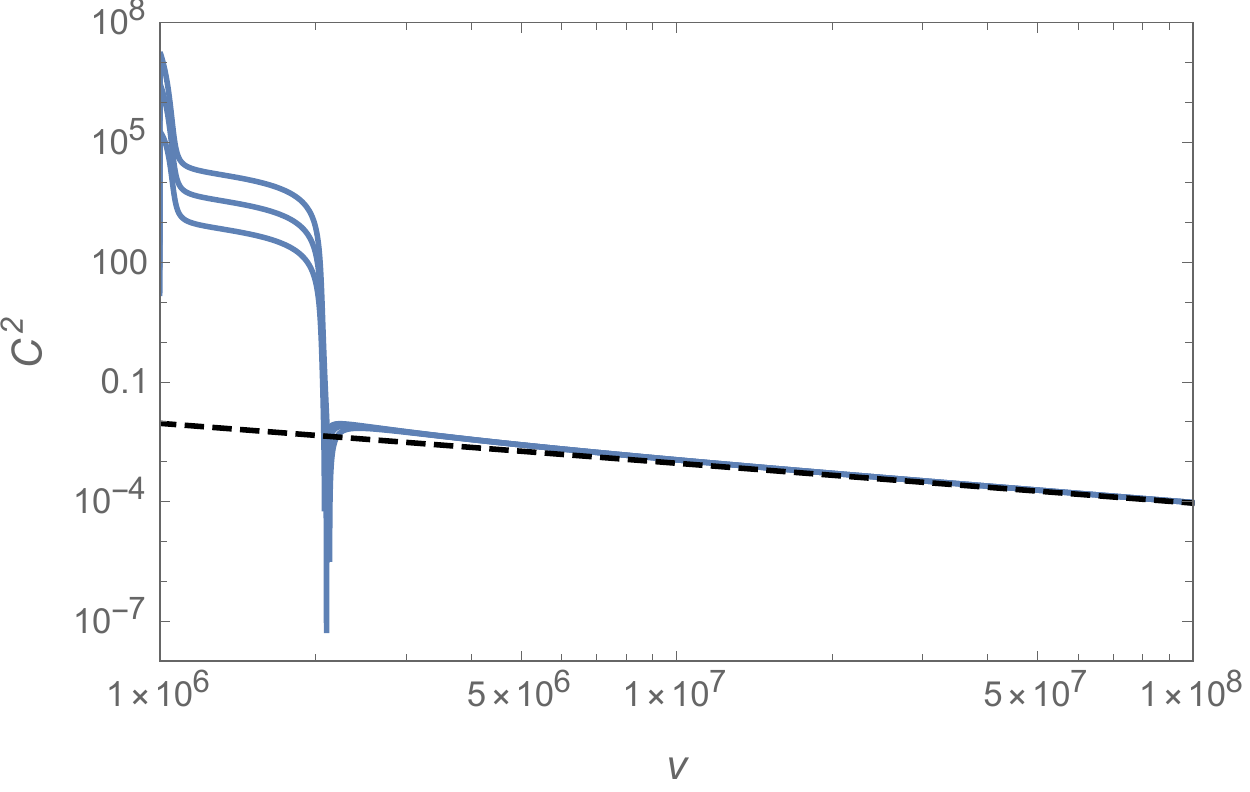} 
	\caption{Illustration of the mass-inflation effect for the Reissner-Nordstr{\"o}m black hole including the Hawking flux in the background geometry. Initial conditions are imposed at early times $v=1$ (left column) and late times $v=10^6$ (right column). The top, middle, and bottom row show the time evolution of $m_+(v)$, $K|_\Sigma$ and $C^2|_\Sigma$, respectively. The early time dynamics depicts the evolution of $m_+(v)$ evaluated for the shell moving along $R(v)$ generated from the initial condition $R(1)=7.5$ for varying initial conditions for $m_+$. Notably, $m_+(v)$ first moves along a plateau before exhibiting a rapid increase when the shell impacts on the apparent Cauchy horizon. This process induces a singularity in $m_+(v)$ situated at approximately $v \approx 14.5$. The divergence of $m_+$ induces a divergence in the curvature scalars when the shell impacts on the apparent Cauchy horizon. The dynamics following from initial conditions imposed at late times differs drastically: in this case $m_+(v)$ reaches the attractor regime \eqref{RN.mass}, indicated by the dashed line. As a result $K|_\Sigma$ remains constant and $C^2|_\Sigma$ vanishes at asymptotically late times, cf.\ \eqref{RN.curvature}.
	\label{shellpositionRN}}
\end{figure*}

The function $m_+(v)$ obtained from the early-time initial conditions is then displayed in Fig.\ \ref{shellpositionRN}.  
This shows that $m_+(v)$ first approaches a constant plateau whose value depends on the initial conditions imposed for $m_+$. Subsequently, the solution terminates in a singularity when the shell impacts on the apparent Cauchy horizon. As illustrated in the left column of Fig.\ \ref{shellpositionRN}, the divergence of $m_+$ induces a divergence in the curvature scalars. Thus, imposing initial conditions at early times leads to solutions which terminate before reaching the late-time attractor identified in the Frobenius analysis.
%

On the right column of Fig.\ \ref{shellpositionRN}, this picture changes radically since the initial conditions are imposed at late-times within the validity range of the Ori model. In order to illustrate this, we solve the same dynamical system with initial conditions imposed at $v=10^6$. At this point the black hole background and in particular $m_-(v)$ and $R(v)$ are already very close to the attractor regime. As shown in the top-right diagram of Fig.\ \ref{shellpositionRN}, this induces the late-time attractor behavior for $m_+(v)$. As a consequence the curvature scalars shown in the middle-right and bottom-right diagram of the figure remain finite. Thus the Reissner-Nordstr{\"o}m black hole does not develop a dynamical singularity once the effect of Hawking radiation is taken into account.
%

\section{Other regular black hole geometries}
\label{sect.RBH}
We complete our analysis by discussing two additional proposals for regular black holes, the renormalization group (RG)-improved black hole solutions \cite{Bonanno:2000ep} and the Bardeen black hole \cite{bardeen1968non,Ayon-Beato:2000mjt}. In these cases the lapse functions take the explicit form
\be\label{fRG}
f = 1 - \frac{2 m r^2}{r^3 +\omega( 2 r+ 9 m)} \, , \quad
M = \frac{r^3}{r^3 +\omega( 2 r+ 9 m)} \, , 
\ee
and
\be\label{fBardeen}
f = 1 - \frac{2 m r^2}{(r^2+a^2)^{3/2}} \, , \quad 
M = \frac{r^3}{(r^2+a^2)^{3/2}} \, , 
\ee
respectively. In order to simplify the analysis we set $\omega = 1$ and $a=1$. While this leads to significant simplifications when analyzing the structure of the equations, fixing the overall scale of the problem still allows to extract the generic features of the dynamics.
\begin{table*}[t!]
	\renewcommand{\baselinestretch}{1.2}
	\begin{tabular}{cccccc}
		\hline
\multirow{2}{*}{}		& \multirow{2}{*}{$f(r)$} & \multirow{2}{*}{$F(v)$} & \multirow{2}{*}{\qquad $K|_\Sigma^\infty$ \qquad } & \quad \multirow{2}{*}{$P(m_+)$} \quad & \quad early solutions reach  \quad \\ 
& & & &  & late-time attractor
\\ \hline
		Reissner-Nordstr{\"o}m & $\Bigg.  1- \frac{2m}{r} + \frac{e^2}{r^2}$ & $\simeq - \sqrt{\frac{15 \pi \, m_{cr}^3}{v}}$ & $\propto v^0$  & linear & no 
		\\ \hline
		Hayward solution \cite{Hayward:2005gi} & $\Bigg. 1 - \frac{2 m r^2}{r^3 +2 m l^2}$ & $\simeq - \sqrt{ \frac{80 \pi m_{cr}^3}{v} }$ & $\propto v^6$ & quadratic & yes \\[1.2ex]
		RG improved black holes \cite{Bonanno:2000ep} & \quad $\Bigg. 1 - \frac{2 m r^2}{r^3 +\omega( 2 r+ 9 m)}$ \quad & \quad  $\simeq -84.5 \, v^{-1/2}$ \quad & $\propto v^6$ & \quad quadratic \quad &  yes \\[1.2ex]
		Bardeen black hole \cite{bardeen1968non,Ayon-Beato:2000mjt} & $\Bigg. 1 - \frac{2 m r^2}{(r^2+a^2)^{3/2}}$ & $\simeq -19.2 \, v^{-1/2}$ & $\propto v^0$ &  linear & no \\[1.2ex] \hline
	\end{tabular}
	\caption{\label{Tab.2} Properties of the late-time attractor controlling the scaling of the curvature in the inner sector of the shell. The evaluation of the RG-improved black holes and the Bardeen black hole is given for the model parameters $\omega=1$ and $a=1$. The column $K|_\Sigma^\infty$ gives the scaling of the Kretschmann scalar evaluated in the limit $v \rightarrow \infty$. Note that the late-time behavior organizes itself in terms of two distinguished universality classes. For the Hayward and RG-improved geometry $K|_\Sigma \propto v^6$ while for the Reissner-Nordstr{\"o}m and Bardeen geometry the curvature scalar remains finite.}
\end{table*}

The analysis of the Ori model for these spacetimes proceeds completely analogous to the previous sections. In particular, the late-time expansion of the mass-function $m_-(v)$ and the motion of the shell are universal in the sense that the leading and subleading terms do not contain free integration constants. The motion of the shell again follows the behavior shown in Fig.\ \ref{shellposition}. The decisive element in the analysis is then given by the structure of the dynamical equation \eqref{mpluseq} determining the mass-function in the inner sector of the shell. For the RG-improved black hole the resulting first-order equation for $m_+(v)$ is
\be\label{mp.RG}
\dot{m}_+ = \frac{\left(9 m_++R^3+2 R\right) \left(\left(9-2 R^2\right) m_+ + R^3+2R\right)}{\left(9 m_-+R^3+2 R\right) \left(\left(9-2 R^2\right) m_- + R^3+2R\right)} \dot{m}_-
\ee
while for the Bardeen black hole one has
\be\label{mp.Bardeen}
\dot{m}_+ = \frac{\left(R^2+1\right)^{3/2}-2 \, R^2 \, m_+}{\left(R^2+1\right)^{3/2}-2 \, R^2 \, m_-} \, \dot{m}_- \, . 
\ee
Two observations are in order. Firstly, \eqref{mp.RG} gives rise to a polynomial $P(m_+)$ which is quadratic in $m_+$. Moreover, a comparison to the curvature invariants given in Table \ref{Tab.3} shows that the factor $9 m_++R^3+2 R$ is the one appearing in the denominators of $K$ and $C^2$. A consistent solution for $m_+$ then requires that this factor vanishes proportionally to $v^{-1}$. This causes a power-law divergence in $K|_\Sigma$ and $C^2|_\Sigma$   which is qualitatively identical to the one found for the Hayward black hole. Thus the late-time behavior of the two geometries follows the same universality class. 

In the case of the Bardeen black hole a consistent solution of \eqref{mp.Bardeen} again requires that $\left(R^2+1\right)^{3/2}-2 \, R^2 \, m_+ \propto v^{-1}$. The crucial difference in this case is that this factor \emph{does not constitute the denominator in the curvature scalars}. As a result $K|_\Sigma$ and $C^2|_\Sigma$ remain finite in the inner sector of the shell. This is identical to the late-time behavior found the  Reissner-Nordstr{\"o}m black hole. Thus, these two geometries also give rise to the same universal late-time behavior close to the Cauchy horizon. The two geometries come with one crucial difference though: the Reissner-Nordstr{\"o}m geometry hosts a spacetime singularity at its center while the Bardeen geometry satisfies the limiting curvature hypothesis everywhere. Thus the Bardeen black hole constitutes a prototypical example of a regular black hole where the mass-inflation effect does not induces a curvature singularity at asymptotically late times. For convenience, we have summarized the key features of the two universality classes and their representatives in Table \ref{Tab.2}.

\section{Summary and outlook}
\label{sect.6}
 In this work we studied the Ori model for mass-inflation for regular black holes. As a novel feature, our analysis includes the mass-loss of the background geometry due to the emission of Hawking radiation. As a result, the asymptotic state of the geometries is an extremal black hole where the event and Cauchy horizons coincide at asymptotically late times. These configurations come with a finite mass, vanishing surface gravity, and zero Hawking temperature, i.e., they correspond to a cold remnant. 
 
 We demonstrated that the fact that the asymptotic geometry is an extremal black hole has profound consequences for the late-time dynamics of the mass-inflation effect. The function $F(v)$ controlling the dynamics of the mass function in the inner sector of the shell vanishes asymptotically, see Table \ref{Tab.2}. This is at variance with the analysis of the same model on a static, non-critical background \cite{Bonanno:2020fgp} where $F(v)$ asymptotes to a constant proportional to the surface gravity at the Cauchy horizon.  As a consequence of this modification, we discover two classes of universal late-time behavior. In the first class curvature scalars including the Kretschmann invariant exhibit a power-law instead of an exponential growth. This class is realized by the Hayward and RG-improved black holes. In the second class the curvature scalars remain finite at asymptotically late times. This occurs for the Reissner-Nordstr{\"o}m solution and Bardeen-type black holes. The fact that the final configuration is an extremal black hole furthermore suggests that radial geodesics do not encounter these singularities in a finite proper time.

A crucial question raised in \cite{Carballo-Rubio:2021bpr,DiFilippo:2022qkl}, is whether the dynamical system can actually reach this salient late-time attractor. Our numerical analysis indicates that the dynamics of the shell and the Hawking effect operate on different time-scales. The impact of the shell onto the apparent Cauchy horizon occurs on much shorter time-scales as the evaporation of the black hole. This feature is universal and does not depend on details of the black hole geometry, cf.\ Fig.\ \ref{shellposition}. Imposing the perturbation at late times, where we expect that the Ori model gives a valid description, we demonstrated that the salient late-time attractor quenching the mass-inflation singularity is reached in all geometries investigated in this work. These include the Hayward black hole, RG improved black hole, the Reissner-Nordstr{\"o}m geometry, and Bardeen-type black holes.

For future investigations, it would be interesting to understand whether there is a mechanism taming the growth of the mass function triggered by the shell impacting on the Cauchy horizon. Refs.\ \cite{Carballo-Rubio:2022kad,Franzin:2022wai} proposed to achieve this by converting the Cauchy horizon to a degenerate horizon with vanishing surface gravity akin to the horizon structure appearing in the final states of the dynamical geometries studied in this work. It would be interesting to actually identify a theory of modified gravity which naturally gives rise to such configurations. One of the best-explored phase spaces, comprising black hole type solutions in quadratic gravity, does not support such features \cite{Quad-Alfio,Quad-Alfio1,BH-QG,BH-QG1,BH-QG2}. 

\bigskip
\acknowledgments
\emph{Acknowledgements} 
 We thank I.\ van der Pas for participating in the early stages of this project. A.-P.\ K.\ thanks the Osservatorio Astrofisico di Catania for hospitality during the initial stages of this project. A.-P.\ K.\ acknowledges financial support by an ACRI Young Investigators Training Program. The work of F.S. is supported by the NWA-grant ``The Dutch Black Hole Consortium''. 
\medskip

\begin{appendix}
\section{Curvature invariants}
\label{app.A}
In order to determine the physics consequences of perturbations on the geometry one should study the dynamics of physical observables. Quantities that naturally lend themselves to such a study are scalar quantities constructed from contractions of curvature tensors as, e.g., the Kretschmann scalar $K \equiv R_{\mu\nu\rho\sigma} R^{\mu\nu\rho\sigma}$ and the square of the Weyl tensor $C^2 \equiv C_{\mu\nu\rho\sigma} C^{\mu\nu\rho\sigma}$. In four spacetime-dimensions these quantities are related by the identity
\be\label{RiemannToWeyl}
C^2 = K - 2 R_{\mu\nu} R^{\mu\nu} + 1/3 R^2 \, . 
\ee
For vacuum solutions of general relativity, satisfying $R_{\mu\nu} = 0$, $C^2$ and $K$ agree. Since the Reissner-Nordstr{\"o}m solution and regular black hole solutions are not Ricci flat, these quantities can be different and in principle also exhibit different asymptotic scaling behaviors. 

The metrics considered in this work take the form
\be\label{met.ans}
ds^2 = - f(r,v) dv^2 + 2 dr dv + r^2 d\Omega^2 \, , 
\ee
where it is convenient to express the lapse function $f(r,v)$ in terms of the Misner-Sharp mass $M(r,v)$ through
\be\label{def.lapse}
f(r,v) = 1 - \frac{2M(r,v)}{r} \, .
\ee
The curvature scalars obtained from \eqref{met.ans} then take the form
\be\label{Kev}
\begin{split}
K = & \, \frac{48 M^2}{r^6} - \frac{64 M M^\prime}{r^5} + \frac{32 (M^\prime)^2}{r^4} \\
& \, + \frac{16 M M^{\prime\prime}}{r^4} - \frac{16 M^\prime M^{\prime\prime}}{r^3} + \frac{4 (M^{\prime\prime})^2}{r^2} \, , 
\end{split}
\ee
and
\be\label{C2ev}
C^2 = \frac{4 \left(r^2 M^{\prime\prime}-4 r M^{\prime} +6 M\right)^2}{3 r^6} \, . 
\ee
Here the prime indicates derivatives with respect to $r$. 

It is instructive to evaluate these curvature scalars for the (regular) black hole geometries discussed in the main part of this work. The results are compiled in Table \ref{Tab.3}. 
\begin{table*}[t!]
	\renewcommand{\baselinestretch}{1.2}
	\begin{tabular}{cccc}
		\hline
	geometry	& $f(r)$ & $K$ &  $C^2$ \\ \hline
	\;	Reissner-Nordstr{\"o}m \; & eq.\ \eqref{lineRN} & $\Bigg. \frac{8 \left(6 m^2 r^2-12 m Q^2 r+7 Q^4\right)}{r^8} $ & $\frac{48 \left(Q^2-m r\right)^2}{r^8}$ 
		\\ \hline
		Hayward  & eq.\ \eqref{eq.hay} & $\; \; \Bigg. \frac{48 m^2 \left(32 l^8 m^4-16 l^6 m^3 r^3+72 l^4 m^2 r^6-8 l^2 m r^9+r^{12}\right)}{\left(2 l^2 m+r^3\right)^6} \; \; $ &  $\frac{48 m^2 r^6 \left(r^3-4 l^2 m\right)^2}{\left(2 l^2 m+r^3\right)^6}$  \\[1.2ex]
		RG-improved &  eq.\ \eqref{fRG} & \quad $\Bigg. \frac{16 m^2 \left(3 r^{12} + k_1 \omega + k_2 \omega^2 + k_3 \omega^3 + k_4 \omega^4 \right)}{\left(9 m \omega +r^3+2 r \omega \right)^6} $ \quad &  $\frac{16 m^2 r^2 \left(54 m r^2 \omega +18 m \omega ^2-3 r^5+2 r^3 \omega \right)^2}{3 \left(9 m \omega +r^3+2 r \omega \right)^6}$  \\[1.2ex]
		Bardeen  & eq.\ \eqref{fBardeen} & $\Bigg. \frac{12 m^2 \left(8 a^8-4 a^6 r^2+47 a^4 r^4-12 a^2 r^6+4 r^8\right)}{\left(a^2+r^2\right)^7}$ & $\frac{12 m^2 r^4 \left(3 a^2-2 r^2\right)^2}{\left(a^2+r^2\right)^7}$  \\[1.2ex] \hline
	\end{tabular}
	\caption{\label{Tab.3} Specific form of the curvature scalars \eqref{Kev} and \eqref{C2ev} evaluated for the Reissner-Nordstr{\"o}m black hole (top line) and the three regular black hole geometries. The explicit form of the line element is given in the equations indicated in the second column. The coefficients $k_i$ parameterizing the curvature for the RG-improved black hole are given in eq.\ \eqref{kdefeq}.}
\end{table*}
The coefficients $k_i$ appearing in the Kretschmann scalar for the RG-improved black hole are
\be\label{kdefeq}
\begin{split}
k_1 = & -4 r^9 (27 m+r) \, , \\
k_2 = & 2 r^6 \left(2187 m^2+432 m r+26 r^2\right) \, , \\
k_3 = & 2 r^3 (-2187 m^3 + 972 m^2 r + 252 m r^2 + 16 r^3) \, , \\
k_4 = & 2 \Big(19683 m^4+11664 m^3 r+ \\ & \qquad 3078 m^2 r^2+360 m r^3+16 r^4 \Big) \, . 
\end{split}
\ee
We observe that for the Reissner-Nordstr{\"o}m geometry the curvature scalars diverge at $r=0$ owed to the vanishing of the denominator at this point. This signals the presence of a curvature singularity in the geometry. Regular black holes change the structure of this denominator in such a way that $K$ and $C^2$ are bounded everywhere as long as $m > 0$ and the model parameters are chosen properly. If $m < 0$ the denominators appearing in the Hayward and RG-improved cases may vanish though. It is this mechanism which underlies the powerlaw divergences of $K|_\Sigma^\infty$ and $C^2|_\Sigma^\infty$ reported in Table \ref{Tab.2}.

\begin{figure*}[t!]
	\includegraphics[width=0.45\textwidth]{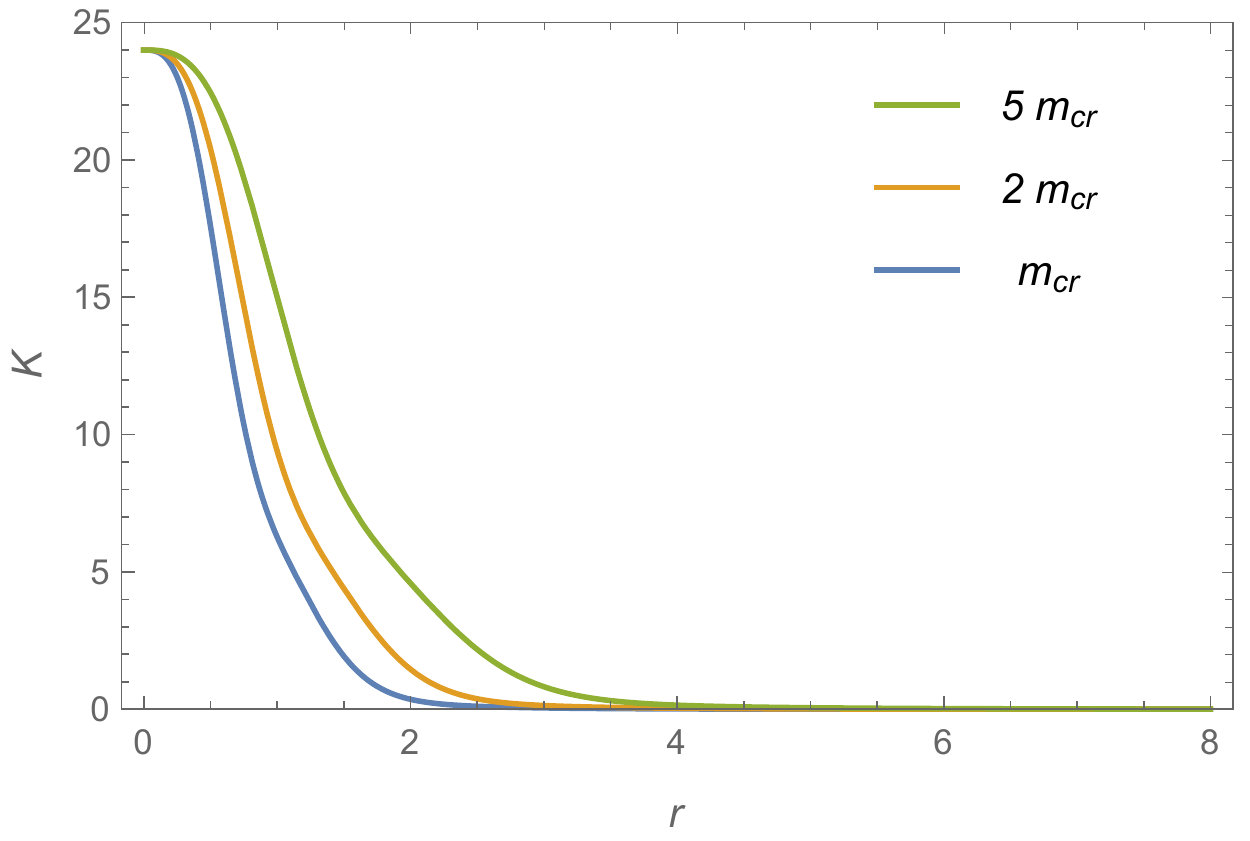}
	\includegraphics[width=0.45\textwidth]{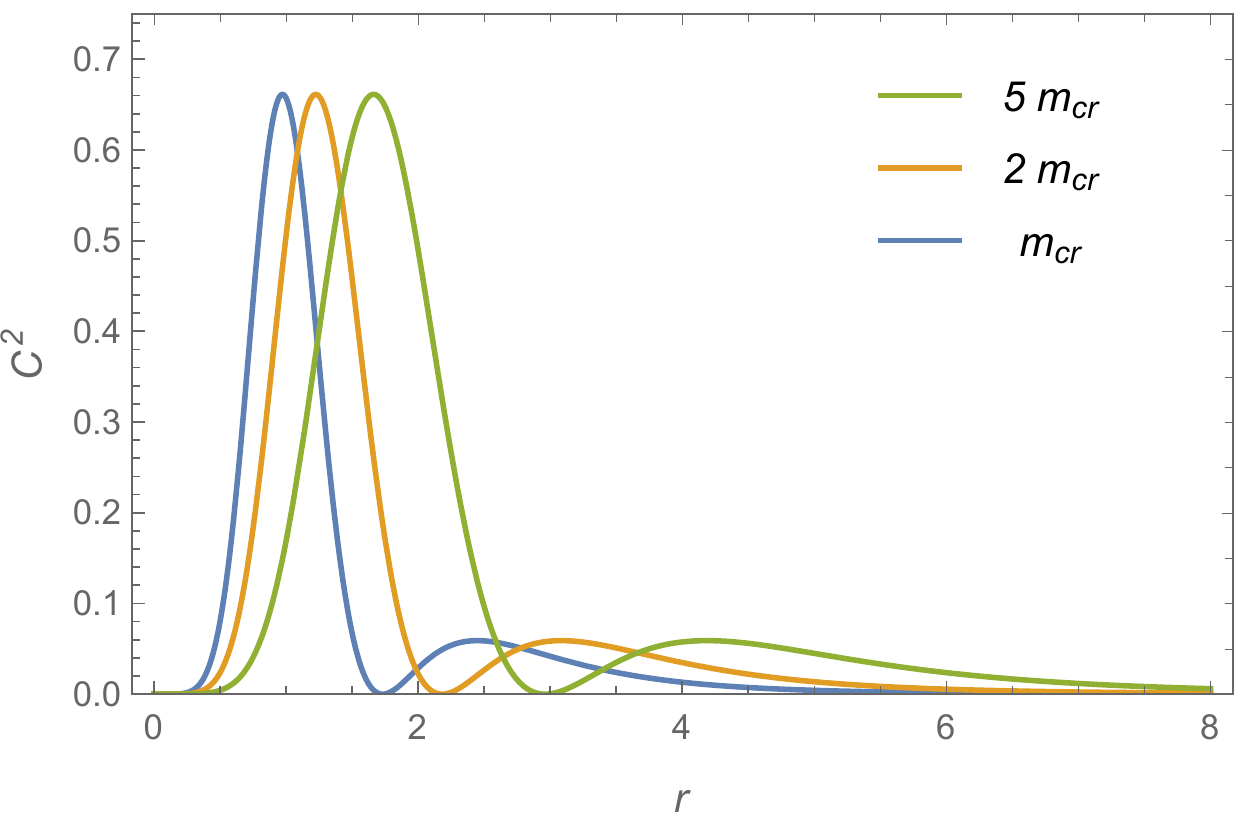}
	\caption{Curvature scalars \eqref{Kev} (top) and \eqref{C2ev} (bottom) evaluated for the Hayward geometry \eqref{eq.hay} with $l=1$.\label{Fig.curvscalars}}
\end{figure*} 
The $r$-dependence of $K$ and $C^2$ obtained for the Hayward geometry \eqref{eq.hay} is shown in Fig.\ \ref{Fig.curvscalars}.
This illustrates the generic feature that $K$ and $C^2$ do not agree for a regular black hole geometry due to extra contributions in Einstein's equations. While $K$ increases monotonically, reaching a finite value at $r=0$, the $C^2$ invariant exhibits zeros at $r=0$ as well as one specific point located between the event and Cauchy horizons.

\begin{widetext}
\section{Frobenius analysis of the late-time attractor}
\label{App.B}
This appendix collects the technical details and subleading terms required in the consistent Frobenius analysis of the late-time attractor. The results for the Hayward black hole are discussed in detail in the text. The analogous formulas for the Reissner-Nordstr{\"o}m geometry (RN), RG-improved black holes (RG), and the Bardeen black hole (B) are are obtained along the same lines and summarized in Table \ref{Tab.4}. 

The line-element specifying the Hayward geometry is given in eq.\ \eqref{eq.hay}. The Frobenius analysis for the mass function $m_-(v)$ in the outer sector of the shell, taking into account the mass-loss from Hawking radiation gives
\be\label{HW.mmasymfull}
m_-(v) \simeq m_{cr} +\frac{480 \pi \,  m_{cr}^4}{v} + 5120 \, m_{cr}^4 \, \left( \frac{5 \pi \, m_{cr}}{v} \right)^{3/2} + \frac{m_{cr}^7}{v^2} \left(\tilde{m} + \text{non-analytic} \right) \, .
\ee
Here the constant $\tilde{m}$ is not determined by the expansion. The occurrence of such a free parameter is expected, since we are approximating the solution to a first order differential equation which comes with one integration constant fixing the solution. It is then remarkable that the first three terms in the expansion \eqref{HW.mmasymfull} are universal in the sense that they are independent of these initial conditions. We also observe that the Frobenius expansion at $\cO(v^{-2})$ requires the contributions of non-analytic terms in order to ensure the cancellation of contributions appearing at $\cO(v^{-3})$ in the expansion. Tracking these terms is beyond the scope of this work. While they may lead to logarithmic corrections to the scaling laws, they will not modify the power-law behavior of the quantities analyzed in this section. 

In the next step, we analyze the late-time behavior of $R(v)$. The numerical analysis indicates that the shell approaches $r_{cr}$ from below, i.e., from radii $r < r_{cr}$. We then select the corresponding sign in the leading term of the large-$v$ expansion. This sign propagates also into the subleading coefficients, so that this branch of solutions has the expansion
\be\label{HW.Rasymfull}
R(v) \simeq r_{cr} -3 \, r_{cr} \, \left( \frac{15 \pi \, r_{cr}^3}{v} \right)^{1/2} - r_{cr}^2 \, \frac{1+585 \pi  r_{cr}^2}{2 v}
+ \frac{r_{cr}^{3/2} \left(27 \left(11686400 \pi ^2-3 \tilde{m}\right) r_{cr}^4+30720 \pi  r_{cr}^2-512\right)}{4096 \sqrt{15 \pi } \, v^{3/2}} \, . 
\ee
In order to ensure that the hierarchy of equations is solved consistently without contributions of higher-order terms missing in the expansion, we included the next term in \eqref{HW.Rasymfull} and verified that it does not enter into the equations determining the coefficients in the expansion.

The next task is to determine the scaling of $m_+$ from eq.\ \eqref{att1}. Here, we use the asymptotic form
\be\label{HW.mpans}
m_+(v) \simeq - 2 m_{cr} + \frac{m_1}{\sqrt{v}} + \frac{m_2}{v} \, .
\ee
In order to deterime the free coefficients $m_1$ and $m_2$, two auxiliary considerations are in order. First, we use \eqref{HW.mmasymfull} and \eqref{HW.Rasymfull} to determine the scaling of the auxiliary function $F(v)$
\be\label{HW.Fv}
F(v) \simeq - 4 m_{cr} \, \left( \frac{5 \pi m_{cr}}{v} \right)^{1/2} \, .
\ee
Note that in this case, it actually suffices to retain the leading term in the expansion. The second observation is that for the late-time attractor \eqref{mplusin2} we must have
\be\label{eqB5}
R^3 + 2 l^2 m_+(v) \simeq \frac{d_1}{v} \, ,
\ee
in order to have a consistent power-law solution compatible with the parameterization \eqref{HW.mpans}. This fixes $m_1 = 48 \sqrt{5 \pi \, m_{cr}^5}$. With this knowledge, we can return to the full equation \eqref{att1} and solve the first non-trivial order to fix $m_2$. This results in the asymptotic expansion
\be
m_+(v) \simeq -2 m_{cr} + 48 \sqrt{\frac{ 5 \pi m_{cr}^5}{v}}
+\frac{2240 \pi  m_{cr}^4+8 m_{cr}^2}{v} \, . 
\ee
As an important corrollary, we note that this implies that
\be\label{HW.d1res}
d_1 = \frac{128 \, m_{cr}^4}{27 v}\, .
\ee
Comparing eq.\ \eqref{eqB5} with the definition of the Misner-Sharp mass in the inner sector of the shell, eq.\ \eqref{eq.hay}, and the results for the curvature scalars given in Table \ref{Tab.3}, we note that it is actually the coefficient $d_1$ which sets the asymptotic scaling behavior of these quantities. Remarkably, \eqref{HW.d1res} is universal in the sense that it does not depend on initial conditions or any free parameters related to the dynamics of the solution. This establishes that the late-time behavior of the perturbed geometry is actually universal. 

The Frobenius analysis for the Reissner-Nordstr{\"o}m geometry literally proceeds along the same lines. In principle, the method is also applicable to the other geometries discussed in the main sections. In these cases the analysis is complicated by a proliferation of square-roots originating from the analytic expressions for the positions of the horizons. Following the strategy of determining the Frobenius coefficients exactly for arbitrary model coefficients $\omega$ and $a$ leads to a significant increase in numerical complexity when simplifying expressions. In order to by-pass this obstacle we then fix $\omega=1$ and $a=1$ and convert the analytic expression to high-precision floating numbers (tracking 50 digits). Based on the insights from the Hayward case, this allows to track the cancellation of leading and also subleading terms in the late-time expansion. The intermediate results leading to the late-time attractor described in Table \ref{Tab.2} are collected in Table \ref{Tab.4} for completeness. 
\begin{table*}[t!]
	\renewcommand{\baselinestretch}{1.2}
	\begin{tabular}{lll}
		\hline \hline
\multirow{3}{*}{$m_-(v)$ \qquad} & RN & $\Bigg. 
m_{cr}
+\frac{30 \pi  m_{cr}^4}{v}
+\frac{720 \pi ^{3/2} \sqrt{15} m_{cr}^{11/2}}{v^{3/2}}
+ \left( \tilde{m} + \mbox{non-analytic} \right) \frac{m_{cr}^7}{v^2} $ \\
                          & RG $(\omega=1) \qquad$  & $\Bigg. m_{cr} + 1.35 \times 10^{3} \, v^{-1} + 1.60 \times 10^8 \, v^{-3/2} + \left( \tilde{m} + \mbox{non-analytic} \right) \, v^{-2}$\\
                          & B $(a=1)$  & $\Bigg. m_{cr} + 2.86 \times 10^3 \, v^{-1} +9.51 \times 10^5 \, v^{-3/2} + \left(\tilde{m} + \mbox{non-analytic} \right) \, v^{-2}$ \\ \hline
\multirow{3}{*}{$R(v)$} & RN & $\Bigg. r_{cr} 
-\frac{2 \sqrt{15 \pi } r_{cr}^{5/2}}{\sqrt{v}}
-\frac{\left(660 \pi  r_{cr}^2+1\right) r_{cr}^2}{2 v} 
-  \frac{1}{4096} \left(  81 \tilde{m} \, r_{cr}^4 - 315532800 \pi ^2 \, r_{cr}^4-30720 \pi  \, r_{cr}^2+512 \right) \left( \frac{r_{cr}^3}{15 \pi v^3} \right)^{1/2}$ \\
      & RG $(\omega=1)$ & $\Big. r_{cr} - 7.35 \times 10^2 \, v^{-1/2} - 2.95 \times 10^5 \, v^{-1} + \left(2.83\times 10^8-2.72 \times 10^{-3} \tilde{m} \right) v^{-3/2}$\\
                          & B $(a=1)$ & $\Bigg. r_{cr} - 81.3 v^{-1/2} - 1.06 \times 10^4 \, v^{-1} + \left(2.02 \times 10^6 - 1.42 \times 10^{-2} \, \tilde{m} \right) \, v^{-3/2}$ \\ \hline
\multirow{3}{*}{$F(v)$} & RN & $\Bigg. -\sqrt{\frac{15 \pi \, m_{cr}^3}{v}}$ \\
                          & RG $(\omega=1)$ & $\Big. -84.5 \, v^{-1/2}$ \\
                          & B $(a=1)$ & $\Bigg. -19.2 \, v^{-1/2}$ \\ \hline
\multirow{3}{*}{$m_+(v)$} & RN & $\Bigg. m_{cr} +\frac{30 \pi  m_{cr}^4}{v} + \frac{720 \sqrt{15} \, \pi^{3/2} \,   m_{cr}^{11/2}}{v^{3/2}}$ \\
                          & RG $(\omega=1)$ & $\Bigg. -8.04 + 4.11 \times 10^3 \, v^{-1/2} + 9.18 \times 10^5 \, v^{-1} $ \\
                          & B $(a=1)$ & $\Bigg. 1.30 + 2.86 \times 10^3 \, v^{-1} $ \\ \hline
\multirow{3}{*}{$M_+(v)|_\Sigma$} & RN & $\Bigg. \frac{m_{cr}}{2}$ \\
                          & RG $(\omega=1)$ & $\Bigg. -1.18 \, v $ \\
                          & B $(a=1)$ & $\Bigg. 5.20$\\ \hline
\multirow{3}{*}{$K|_\Sigma$} & RN & $\Bigg. \frac{8}{m_{cr}^4}$ \\
                          & RG $(\omega=1)$ & $\Bigg. 2.40 \times 10^{-4} \, v^6$ \\
                          & B $(a=1)$ & $\Bigg. 1.44$\\ \hline
\multirow{3}{*}{$\Bigg. C^2|_\Sigma$} & RN & $\Bigg. \frac{2880 \pi}{m_{cr} \, v}$ \\
                          & RG $(\omega=1)$ & $\Bigg. 8.00 \times 10^{-5} \, v^6$ \\
                          & B $(a=1)$ & $\Bigg. 3.70 \times 10^{-2}$
 \\[1.2ex] \hline \hline
	\end{tabular}
	\caption{\label{Tab.4} Late-time expansion of the functions controlling the dynamics of the Ori-model in the inner sector of the shell. The expansion order of each function is sufficient to arrive at the leading terms in the scaling behavior of the Misner-Sharp mass $M_+(v)|_\Sigma$ and the curvature scalars $K$ and $C^2$ evaluated at the position of the shell.}
\end{table*}
%
\end{widetext}
\section{Geodesic equations}
\label{app.C}
In order to determine the geodesic structure of spacetime, we consider the motion of radially infalling observers and compute the relation between the coordinate $v$ and the observer's proper time $\tau$. Starting from the line element \eqref{met.ans}, the $v$-component of the geodesic equation is
\be\label{eq.geo1}
\ddot{v}=-\frac{1}{2}\frac{\partial f}{\partial r}\dot{v}^2 \, , 
\ee
where the dot represents a derivative with respect to proper time. The normalization for the four-velocity of the radial observer furthermore supplies the relation
\begin{equation}\label{Leq}
    L=\frac{1}{2} ( f \dot{v}^2-2 \dot{r} \dot{v})=\frac{\epsilon}{2} \, . 
\end{equation}
with $\epsilon=0,1$ for lightlike or timelike geodesics, respectively. For a static geometry where $f$ is independent of $v$, %
\be\label{conserved}
   \frac{\partial L}{\partial\dot{v}}= const \, , \\
\ee
is conserved along geodesic motion.

To understand the implications of the curvature singularities induced by the late-time attractors identified in Table \ref{Tab.2}, we solve these equations for timelike geodesics close to the Cauchy horizon. Considering \eqref{eq.geo1}, for a static, non-critical black hole, we first note that near the inner horizon
\be\label{eq.surfacegravity}
-\frac{1}{2}\frac{\partial f}{\partial r} \simeq \kappa_{-}>0 \, . 
\ee
Eq.\ \eqref{eq.geo1} is then readily solved in this limit, yielding the relation
\be\label{geo.noncritical}
\tau = \frac{1}{\kappa_{-}} e^{-\kappa_{-} v} + const \, . 
\ee
The important insight from this result is that a massive observer can reach the singularity at $v=\infty$ \emph{in finite proper time}. Hence the strength of the singularity becomes an important question when asking whether a geodesic can be continued beyond this point \cite{Bonanno:2020fgp}. 

In the presence of Hawking radiation this analysis is radically altered though. In this case the late-time structure of spacetime is given by an extremal black hole where $\kappa_- =0$. Therefore, eq.\ \eqref{eq.geo1} gives the linear relation 
\be\label{geo.critical}
\tau = c v + const \, .
\ee
with $c$ an integration constant. 
The crucial difference to \eqref{geo.noncritical} is that the timelike geodesic requires \emph{an infinite amount of proper time to reach} $v=\infty$. 

\end{appendix}

\bibliography{bh_bib}

\end{document}